\documentclass{article}[15pt]

\usepackage{amsmath,epsfig,color,calc,rotating}
\usepackage{graphics}
\usepackage{wrapfig}
\usepackage{ulem}

\usepackage[super,sort&compress]{natbib}
\bibpunct{[}{]}{,}{n}{}{,}

\begin{document}

\title{
Volcano Plots in Analyzing Differential Expressions with mRNA Microarrays
\vspace{0.2in}
\author{
Wentian Li \\
{\small \sl The Robert S. Boas Center for Genomics and Human Genetics}\\
{\small \sl The Feinstein Institute for Medical Research,
North Shore LIJ Health System}\\
{\small \sl Manhasset, 350 Community Drive, NY 11030, USA.}\\
}
\date{}
}
\maketitle  
\markboth{\sl Li }{\sl Li}

\vspace{0.3in}

\begin{center}
{\bf ABSTRACT}: 
\end{center}

Volcano plot displays unstandardized signal (e.g. log-fold-change) 
against noise-adjusted/standardized signal (e.g. $t$-statistic or 
$-\log_{10}(p$-value) from the $t$ test). We review the basic 
and an interactive use of the volcano plot, and its crucial role 
in understanding the regularized $t$-statistic.  The joint 
filtering gene selection criterion based on regularized 
statistics has a curved discriminant line in the volcano plot, 
as compared to the two perpendicular lines for the ``double filtering" 
criterion. This review attempts to provide an unifying
framework for discussions on alternative measures of
differential expression, improved methods for estimating
variance, and visual display of a microarray analysis result.
We also discuss the possibility to apply volcano plots to other
fields beyond microarray.

\begin{center}
{\bf KEYWORDS}:
\end{center}

microarray; volcano plot; signal-to-noise ratio; regularization

\large

\section{Introduction}
\indent

The microarray technology allows simultaneous measurements of
messenger RNA level of thousands of genes,  and its adoption dramatic
changes the way biological and biomedical research 
is carried out  \citep{schena,young,butte,slonim,stoughton,trevino,trachtenberg}.
In particular, the more labor-extensive real-time PCR can be
replaced by microarray profiling in a preliminary round, 
as the general agreement between the two methods is considered to be good
\citep{etienne,dallas,morey}. As an emerging technology,
there are still many issues to be worked out, such as the
consistency among different platforms 
\citep{park,larkin,irizarry,draghici,kuo,patterson,chen07,wen,zli12}
as well as their integration \citep{allen}, batch effect 
\citep{churchill,cli,baggerly,kitchen,cchen,gagnon}, level, source, and distribution 
of noise \citep{ioannidis,raser,eindor,maheshri,zeisel,thomas,kitchen11,posekany,tang}, limit of dynamic 
range \citep{sharov}, etc.  However, with better probe design 
\citep{yang,trachtenberg}, better data quality control \citep{shi,shi06,mccall}, 
better data reporting requirement \citep{miame,ioannidis09}, 
better normalization scheme 
\citep{quack,vande,fujita,steinhoff,stafford,astola,onskog,halpert},
and better understanding of the study goals, these are not insurmountable problems.

Analyzing large amount of expression data from microarray
experiments was thought as a major challenge in early days,
but this problem was over-estimated. First, the amount of data
from thousands of genes and a hundred or so samples is still much smaller
than, e.g., the data generated by whole-genome association studies
\citep{estrada}
or next generation sequencing \citep{schadt}, and a moderately sized computer 
might handle the data without problems. Second, no brand-new statistical 
learning methods have to be invented and existing machine learning techniques 
\citep{hastie} could already extract meaningful information from the data.
Third, the problem of larger number of false positives due
to the large number of genes being profiled has been addressed
and properly handled \citep{storey,storey-q,reiner,pawitan,schw}.
Fourth, in using multiple genes in constructing classifiers,
the well known ``large $p$, small $n$" problem (large number of
variables with small number of sample size) can be solved
by the variable/subset/feature/model selection techniques
\citep{xing,liyang,ambroise,guyon,ding,wli06,liao,zhao}

One of the most common applications of microarrays is 
``differential expression" profiling: identifying mRNAs/genes
whose expression level is very different under two conditions,
e.g., with disease and being healthy. Not only could differentially 
expressed genes provide insight into the biological processes
involved in disease etiology,  but also these can be used 
as biomarkers for diagnosis \citep{golub,hedenfalk,dhan,adib,yeatman}
or prognosis \citep{pomeroy,vij,colman,kim}. 
The phrase ``differential expression" means that the {\sl averaged} expression
level of a mRNA/gene in one phenotype-specific group is much
{\sl larger} or {\sl smaller} than that in another group. However, the
terms ``{\sl average}" and ``{\sl larger/smaller}" are up to various
interpretations.

There are at least two definitions of average: arithmetic mean
($E[x]= \frac{1}{n} \sum_{i=1}^n x_i$)
or geometric mean ($G[x] = (x_1 x_2 \cdots x_n)^{1/n}$). 
For fluorescence-light-intensity based microarray data $x$, 
it is a common practice to logarithmically transform the 
data $x'=\log_{10}(x)$, because $x'$ fits better than $x$ to a 
normal distribution (without losing generality, the base
of the logarithmic function is chosen at 10 in this review). 
Then $E[x']=  \frac{1}{n} \sum_{i=1}^n \log_{10}(x_i) 
= \log_{10} (x_1 x_2 \cdots x_n)^{1/n}= \log_{10} G[x]$,
connecting the two means. Yet another measure of average
is the median, being unaffected by log-transformation,
which has been used in \citep{jain}. 

Deciding ``how large one group's average is compared to 
the other" is no less trivial. Fold-change and $t$-statistic
are the two main choices for measuring differential expression.
In microarray analysis field, these two measures have been
in and out of favor at various time. Fold-change had been
commonly used before it was pointed out that it did not
take the noise into account \citep{chen97,baldi}. $t$-statistic
enjoyed its acceptance until another round of papers suggesting
that genes selected by fold-change are more consistent among
different microarray platforms than those selected by $t$-statistics
\citep{shi05,shi06,guo}. This result led to more discussion
on the relationship between reproducibility and accuracy 
\citep{mzhang,mzhang2,boule},
and between biological and statistical signal \citep{witten07}.

Despite development of sophisticated methods for microarray 
analysis, one question we analysts hear the most from the end-users
is ``should I use fold-change or $t$-statistic?".
The problem with fold-change is that the same fold-change 
value will be less impressive if the variance is large. 
Although $t$-statistic aims at taking the noise level 
into account, the practical problem is that the variance 
may not be estimated reliably, especially when the sample 
size is small. An answer provided by this review is basically 
``use both": the volcano plot is exactly such a 
visual tool to display both fold-change and $t$-statistic.

This review is organized as follows: Section 2 establishes 
a relationship between the fold-change and $t$-statistic; 
Section 3 introduces volcano plots and its basic usage;
Section 4 summarizes the idea of ``moderated", ``regularized", 
``penalized" statistics by adding an extra positive term 
to the sample-based variance or standard error;
Section 5 discusses the regularized statistics in the
context of volcano plot; Section 6 surveys the 
software packages in Bioconductor that are relevant to 
this review; Section 7 introduces the idea of stratified 
volcano plots; and the final Section is the discussion 
and conclusion section.

All plots in this paper use the same published dataset
containing 37 case/patient samples and 18 control samples,
with 48804 probesets in Illumina platform, normalized
by ``quantile normalization".

\begin{figure}[ht]
 \begin{center}
 \begin{turn}{-90}
 \epsfig{file=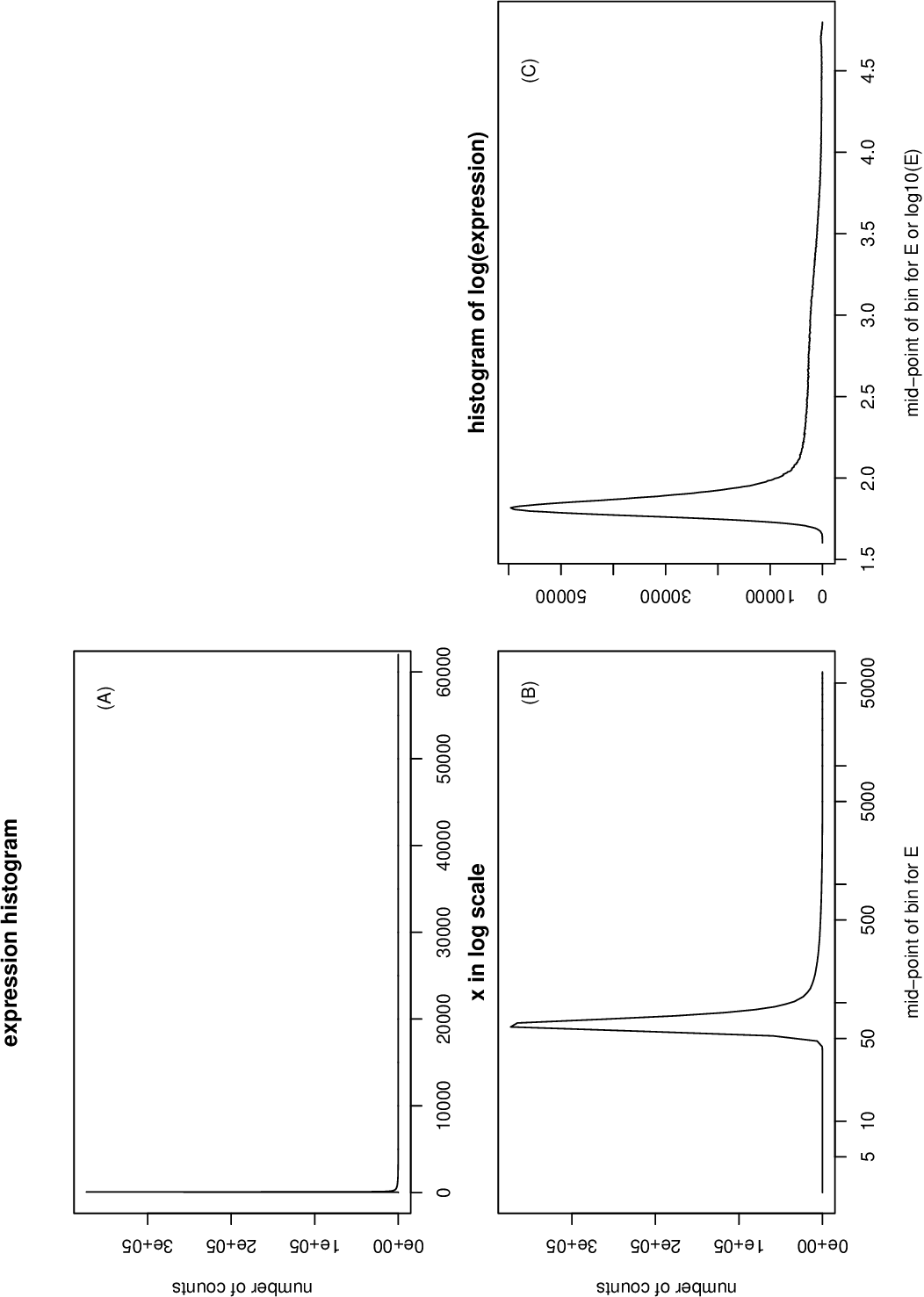, height=15cm}
 \end{turn}
 \end{center}
\caption{
\label{fig1}
Histogram of expression levels of a microarray experiment:
(A) in linear scale.
(B) $x$-axis in a log scale. 
(C) for log-transformed expression.
}
\end{figure}

\begin{figure}[ht]
 \begin{center}
 \begin{turn}{-90}
 \epsfig{file=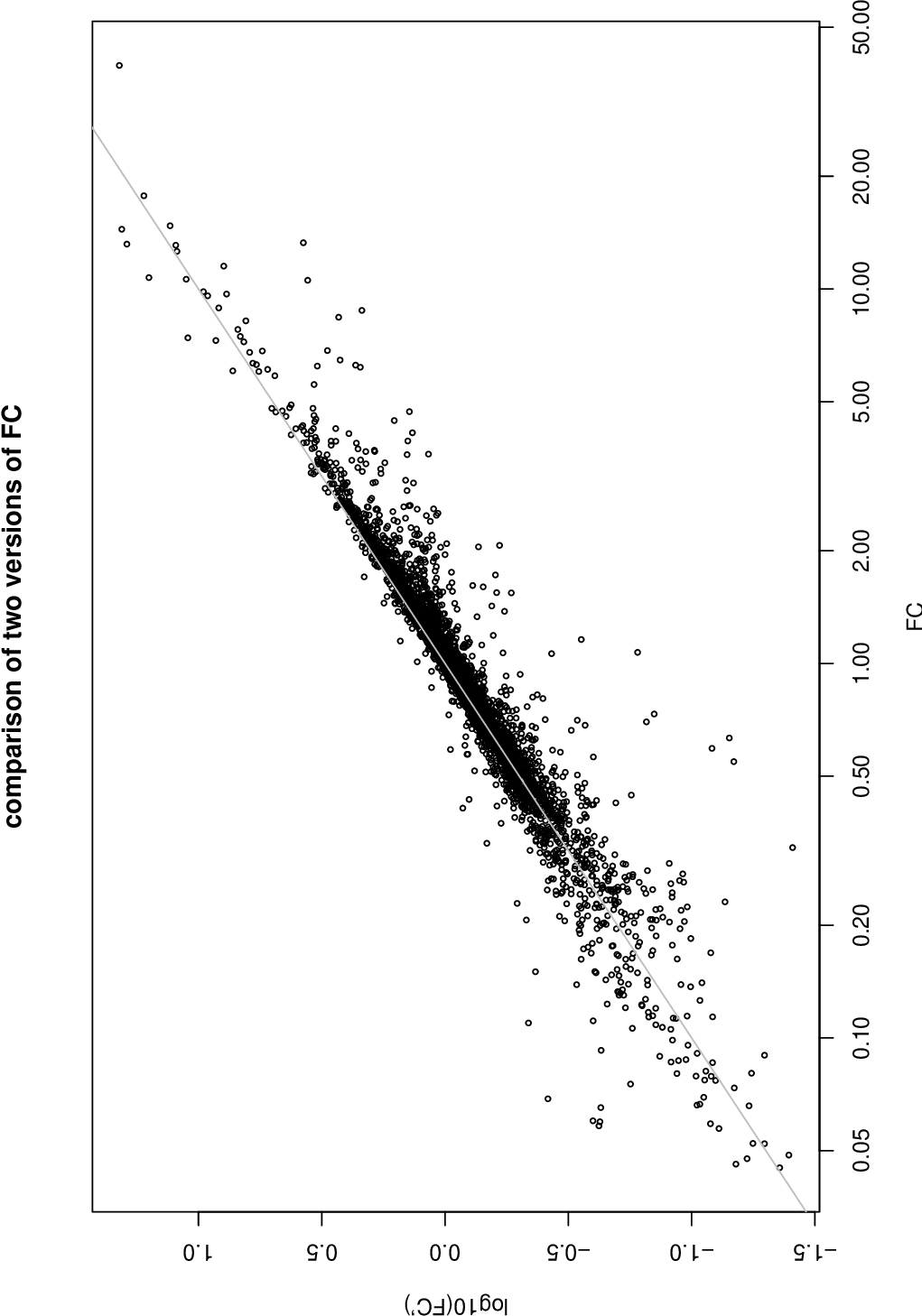, height=13cm}
 \end{turn}
 \end{center}
\caption{
\label{fig2}
Comparison of two definitions of fold-changes. The $x$ is
FC $=\langle E_1\rangle/\langle E_0\rangle$ in log scale. The $y$
is the $\log_{10}(FC')= \langle \log_{10} E_1\rangle - \langle \log_{10} E_0\rangle $ 
(Eq.(\ref{eq-fc2})).
}
\end{figure}

\section{Fold-change and $t$-statistic: signal and signal-to-noise ratio}
\indent

Fold-change (FC) and $t$-statistic seem to be two very different quantities:
one is intuitive and a straightforward measure of differences, another
is rooted deeply in the field of statistics. However, with logarithm transformation
there is a relationship between the two. 

The need for logarithmic transformation can be illustrated by Fig.\ref{fig1}.
Fig.\ref{fig1} shows the three histograms of fluorescence-light intensity
$E$ of a microarray experiment which is indicative of
the number of mRNA copies hybridized to the probe, thus
a measure of mRNA expression level:  (A) in regular scale, (B) in log-transformed
$x$-axis scale, and (C) of $\log_{10}(E)$ itself. Without the logarithmic
transformation, the distribution of $E$ is very long-tailed, and very
skewed (asymmetric). With the log transformation (or other similar
transformations in a recognition that log transformation cannot handle
zero level \citep{ohara}), even though the distribution is still not a perfect 
normal distribution, it is much more ``normal-like". 

There are other advantages of a log transformation, e.g. variance is more stablized
and does not tend to increase with the mean; it is consistent with
a psycho-physics law relating human sensation to the logarithm of
the stimulus level \citep{fechner}. Note that for 
non-fluorescence-light-density-based technologies for measuring 
expression level, such as digital expression and RNA-seq \citep{auer}
we lose this ground for justifying log-transformation.
The decision on whether to use a transformation to become
a normal distribution, or whether to model the data by another
distribution completely, such as the Poisson distribution,
is empirically based on the histogram of the data
\citep{robinson,oshlack,deseq,bayseq,degseq,cumbie,lee11,chen11,tarazona,witten11,kvam,jli12}.
However, we also notice that Poisson distribution is
approximately a normal distribution when its mean is large.

The simplest definition of FC is:
$FC = \langle E_1 \rangle/\langle E_0 \rangle$, where the arithmetic 
average is over the fluorescence-light intensity of samples in 
group 1 (e.g. diseased group) and group 0 (e.g. control group).
The logarithm of FC is:
$\log_{10}(FC)= \log_{10} \langle E_1 \rangle/\langle E_0 \rangle
= \log_{10} \langle E_1 \rangle - \log_{10} \langle E_0 \rangle
\approx \langle \log_{10} E_1 \rangle - \langle \log_{10} E_0 \rangle$.
Reversing the order of averaging and log-transformation operations
usually does not lead to identical values, so the above expression is
only an approximation. We can have a second definition of FC
called FC':
\begin{equation}
\label{eq-fc2}
\log_{10}(FC') = \langle \log_{10} E_1 \rangle - \langle \log_{10} E_0 \rangle
\end{equation}
Fig.\ref{fig2} shows that FC is mostly similar to FC' and
we do not distinguish the two definitions. The same conclusion
is also reached in \citep{witten07}.

The $t$-test is an example of statistical testing whose goal is
to compare any observed result with chance events. The statistic
used in $t$-test (e.g. \citep{snedecor}) is the difference of
arithmetic means in two groups divided (``standardized") by
the estimated standard deviation of that difference. Standard  
deviation of parameters (e.g., sample mean, sample variance)
is often called ``standard error" (SE) \citep{snedecor}.  One requirement
for using $t$-test is that values in two groups roughly follow
normal distributions. As discussed above,
we need to log transform the fluorescence light intensity $E$ to have a
normal-like distribution, so $t$-statistic is:
\begin{equation}
\label{eq-t}
 t= \frac{ \langle \log_{10} E_1 \rangle - \langle \log_{10} E_0 \rangle}
{SE_{\langle \log_{10} E_1 \rangle -\langle \log_{10} E_0 \rangle}}
= \frac{ \langle \log_{10} E_1 \rangle - \langle \log_{10} E_0 \rangle}
{ \sqrt{\frac{s_1^2}{n_1} + \frac{s_0^2}{n_0} }}
\end{equation}
where the second formula was due to Welsh \citep{welsh}, 
who assumed different variances in group 1 and group 0 and
provided an estimation of $SE$ ($s_1^2$ and $s_0^2$ are the estimated 
variances (of $\log_{10}(E)$) of group 1 and 0, and $n_1$, $n_0$ are 
number of samples in the two groups).

Comparing Eq.(\ref{eq-fc2}) and Eq.(\ref{eq-t}), we
establish a relationship between $\log_{10}(FC)$ and $t$-statistic:
$t$ is $\log_{10}(FC)$ 
standardized by the noise level as measured by the pooled standard 
error. There are parallel contrasts of measures in other fields, such
as the {\sl signal-to-noise ratio} (vs. signal by itself) in engineering,
{\sl standardized effect size} (vs unstandardized effect size) in
statistical behavioral science, quantitative psychology, epidemiology, 
and meta-analysis \citep{cohenbook}. The exact relationships
between them, however, require more careful examination; for example, $t$-statistic
increases with sample $n$ by the factor of $\sqrt{n}$ when it is not zero,
whereas standardized effect size does not change with the sample size.

\begin{figure}[th]
 \begin{center}
 \begin{turn}{-90}
 \epsfig{file=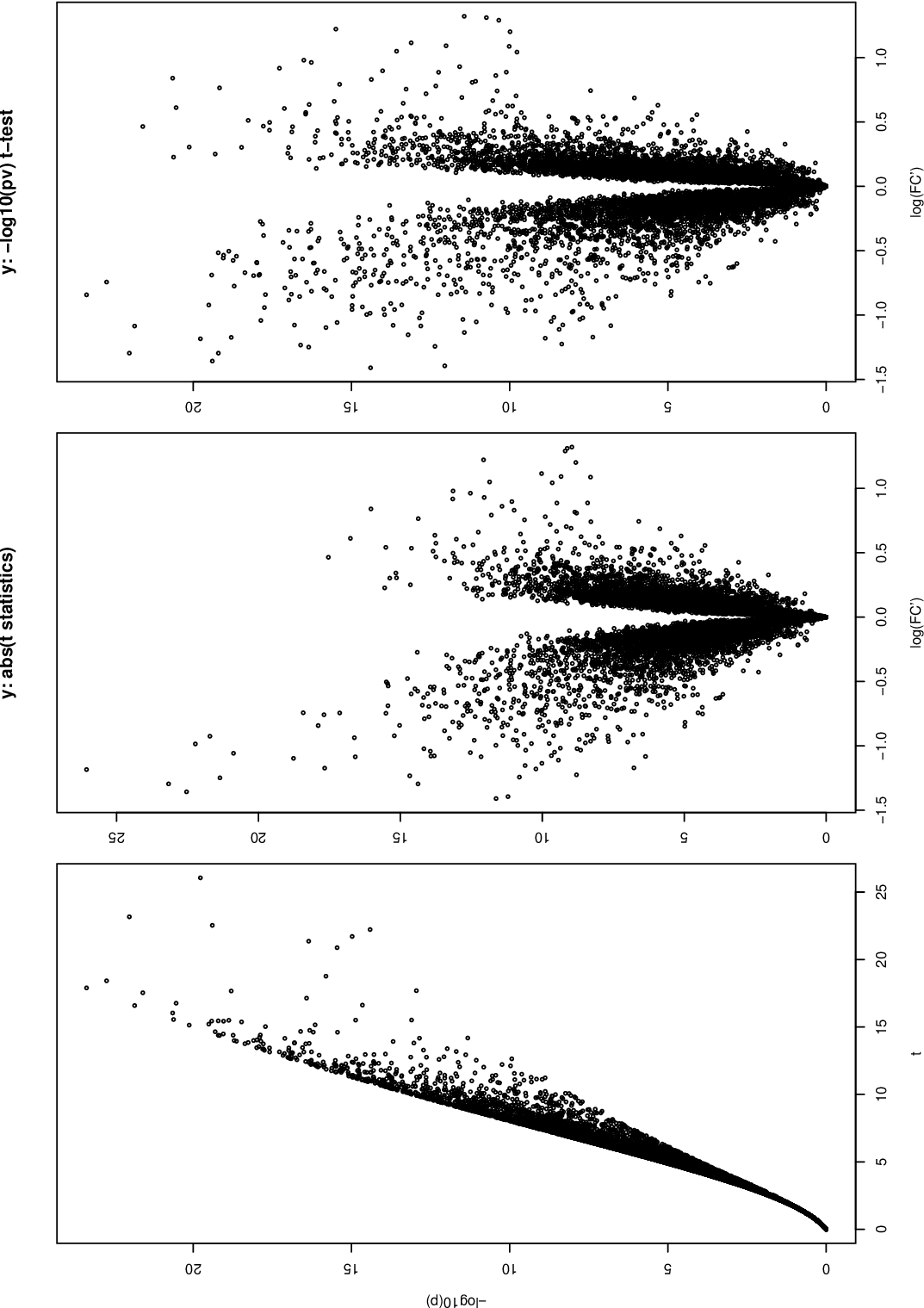, height=13cm}
 \end{turn}
 \end{center}
\caption{
\label{fig3}
(A) $x$-axis: $t$-statistic, $y$-axis: $-\log_{10}(p$-value)
of $t$-test.
(B) Volcano plot using $t$-statistic as the $y$-axis ($x$-axis is
log$_{10}$FC).
(C) Volcano plot using $-\log_{10}(p$-value) as the $y$-axis.
}
\end{figure}

\section{Volcano plot and its basic use}
\indent

If the noise level is known or can be reliably estimated, it is
of course preferable to measure differential expression that
takes the noise level into account, such as $t$-statistic.
In reality, not only is smaller sample sizes an issue
for variance estimation, but also, if systematic error exists,
we may not improve the situation by increasing the sample size.
For example, it is observed that noise level during the hybridization
stage is much higher than that during the sample preparation
or amplification stage \citep{tu}. If a probe sequence for 
an mRNA is highly represented in the genome, cross-hybridization
can be a cause of error and variation, and the probability
of this error does not seem to decrease with large sample sizes.

Facing this reality, we might just display and use both FC and
$t$-statistic, and this is what the volcano plot does. Volcano plot
most often refers to the scatter-plot with $-\log_{10}(p$-value)
from the $t$-test as the $y$-axis and (log$_{10}$)FC as the $x$-axis
\citep{gibson,cui03,alvord}. However, $t$-statistic and $-\log_{10}(p$-value)
are highly correlated (see Fig.\ref{fig3}(A)), and whether
the $t$ (Fig.\ref{fig3}(B)) or $-\log_{10}(p$-value) (Fig.\ref{fig3}(C))
is used in the $y$-axis, the outcome is very similar. 
The reason why $t$ and $p$-value from $t$-test is not
one-to-one corresponding (Fig.\ref{fig3}(A)) is because
in determining $p$-value, Welsh's $t$ distribution has
a degree of freedom parameter which also depends on the data
\citep{pan}.

\begin{figure}[th]
 \begin{center}
 \begin{turn}{-90}
 \epsfig{file=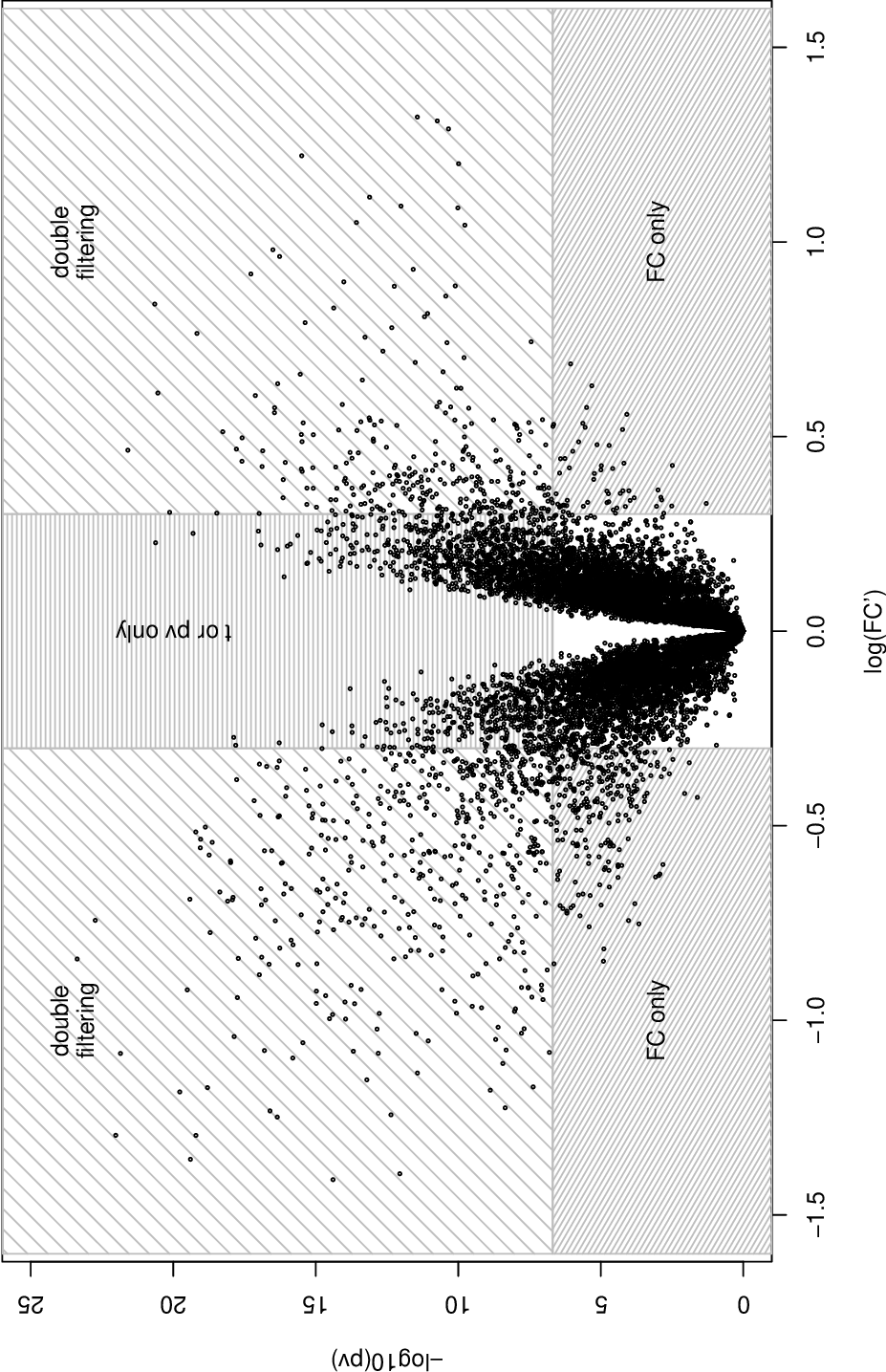, height=13cm}
 \end{turn}
 \end{center}
\caption{
\label{fig4}
Illustration of the double filtering criterion (upper-left and upper-right
corners shaded by sparse lines), FC-only single-gene criterion 
(lower-left and lower-right corners shaded by dense lines), and 
$t$-test-only single-gene criterion (``football goalpost" in the middle
shaded by dense horizontal lines).
}
\end{figure}

\begin{figure}[th]
 \begin{center}
 \begin{turn}{-90}
 \epsfig{file=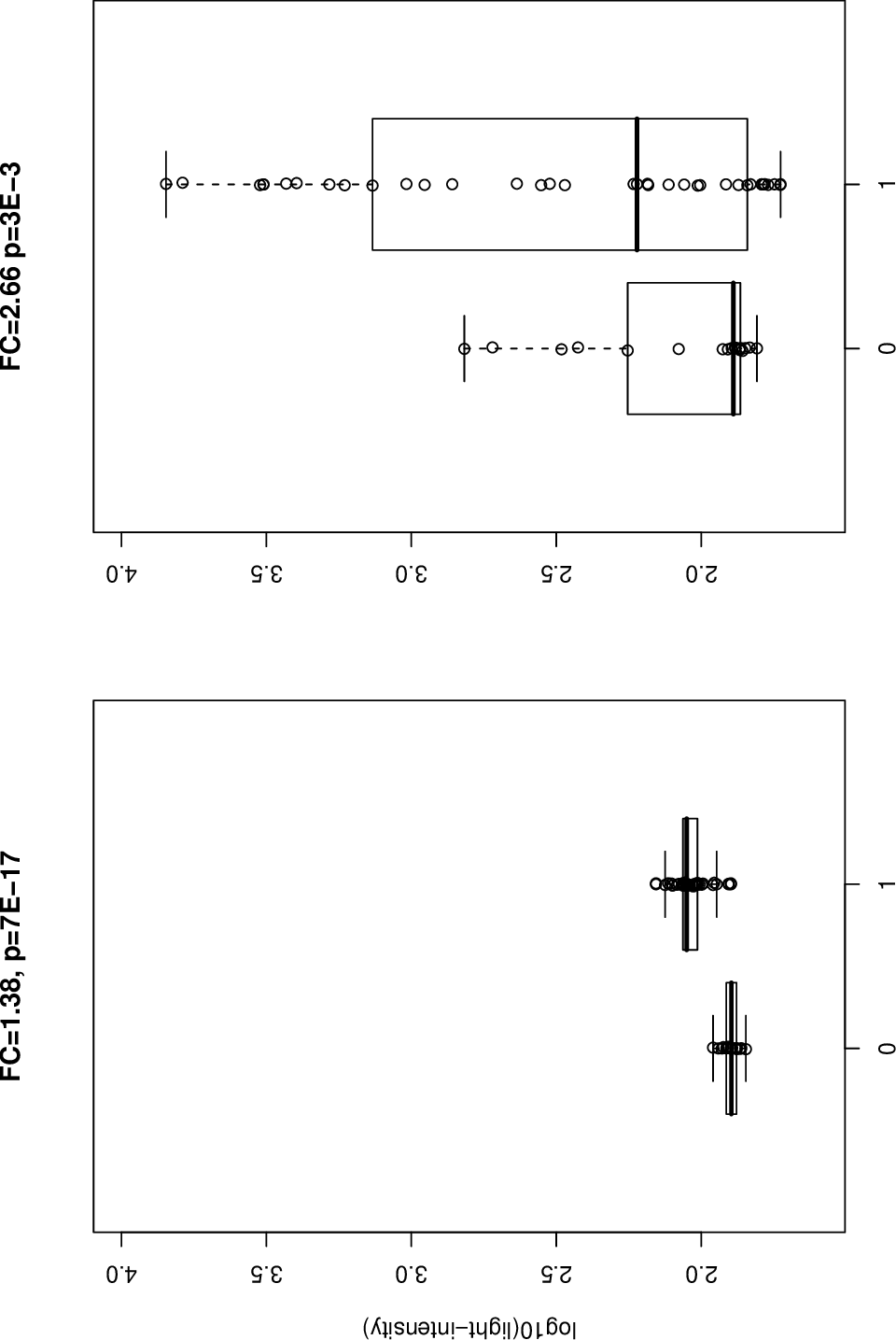, height=11cm}
 \end{turn}
 \end{center}
\caption{
\label{fig5}
(A) a gene with a significant $t$-test result ($p$-value = $7.7 \times 10^{-17}$)
but only moderate fold-change (FC=1.38).
(B) a gene with large fold-change (FC=2.66) but weaker $t$-test
significance ($p$-value= 3 $\times 10^{-3}$).
}
\end{figure}

\begin{figure}[ht]
\begin{center}
\begin{verbatim}
FC <- c(1.03, 2.4, 0.49, 0.6, 1.3, 0.9)
pv <- c(0.5, 3E-7, 2E-4, 5E-3, 0.08, 4E-4)
gname <- c("AAC","ARG1", "CCDC4", "DEFB4", "EIF1", "GNAQ")
x <- log10(FC)
y <- -log10(pv)
plot(x,y)
identify(x,y, n=6, labels=gname)
\end{verbatim}
\end{center}
\caption{\label{Rscript1}	
An {\sl R} script illustrating the use of interactive plotting function
{\sl identify} in volcano plots. Suppose there are 6 genes whose
fold-changes ({\sl FC}) and $t$-test $p$-values ({\sl pv}) are given,
and the gene names are in {\sl gname}. After the logarithmic
transformations, the volcano plot is drawn by {\sl plot(x,y)}.
}
\end{figure}

The basic use of volcano plots is to survey genes that could be 
selected by one differential expression criterion but not the other. 
 The familiar ``double filtering" \citep{zhang} used by many 
groups is to set the gene selection criterion by: (i) 
$|\log_{10}FC|  > \log_{10}FC_0$; and (ii) $t > t_0$. 
Equivalently, it can be defined as (i) $|\log_{10}FC| > \log_{10}FC_0$; 
and (ii) $p-$value $< p_0$. FC$_0$, $t_0$, $p_0$ are preset 
threshold values for fold-change, $t$-statistic, and $t$-test $p$-value. 
The double filtering criterion corresponds to a cutting out
of two rectangular corners away from the origin (Fig.\ref{fig4}).
The single filtering criterion corresponds to delineating 
(away from origin) regions by horizontal and vertical lines.
Then genes chosen by the single {\sl but not} by the
double filtering criterion are in the three disjointed
regions shaded in Fig.\ref{fig4}.

These genes in the shaded area are often not selected for 
good reasons: (i) genes with large fold-change but nevertheless 
insignificant test result may be caused by a few outliers with very 
large values in one group.  (ii) genes with significant test result 
(large $t$'s and small $t$-test $p$-values) but low fold 
change could be false signal due to low variance, which
can be caused by batch effect \citep{leek10}, or low expression
level (to be discussed later).  A volcano plot allows us to 
pick some genes from the shaded regions in Fig.\ref{fig4} for 
further examination.  

To understand better the difference between the two single-gene
filtering criteria (horizontal and vertical lines in Fig.\ref{fig4}),
we show two examples of genes selected by the two
single filtering criteria in Fig.\ref{fig5}. 
Fig.\ref{fig5}(A) is a gene selected by $t$-test $p$-value only 
($p= 7.7 \times 10^{-17}$) while FC is lower than 2 
(FC=1.379). If the true variance is indeed low and we estimated 
it correctly from 17 control samples, then we trust that 
this gene is significantly differentially expressed.

On the other hand, The gene in Fig.\ref{fig5}(B) is selected by FC only (FC=2.66) 
whereas the $p$-value is only $3 \times 10^{-3}$. This gene can
still be a significantly differential-expression if
the large variance in the case group is due to something
else, e.g. sub-disease types. Statistical test alone
should not be the only foundation for selecting potentially
relevant genes, and volcano plot is a way to pick genes
that may not lead to the smallest $p$-values.

Interactively selecting genes in a volcano plot can be done
in the statistical package {\sl R} ({\sl http://www.r-project.org/}).
The {\sl R} function for this purpose is {\sl identify}, which identifies
the closest point in a scatter plot to the position clicked
by the mouse button. Then information about that point can be
printed on screen or in an {\sl R} session window.  Because
volcano plot is usually crowded already, one would prefer to avoid
printing long character strings to the screen -- a gene name
should be often appropriate (human gene names are  
standardized by HUGO gene nomenclature committee:
{\sl http://www.genenames.org/}). An illustrative {\sl R} script 
for using {\sl identify} is included in Fig.\ref{Rscript1}.

Volcano plot does not show the average expression level
of a gene, thought this information can be added using
colors (X Hua, X Yan, S Yancopoulos, Y Yang, W Li, 
``STRAT-VOL: stratified volcano plot for microarray expression
analysis", unpublished draft). Nevertheless, the relative
magnitude of standard deviation of a gene is provided
by the volcano plot, as it is proportional to the tangent of the angle
between the point-to-origin line and the $y$-axis 
(see Section 5 for more details).

\section{Robust variance estimation and regularization}
\indent

The essential difference between FC and $t$-statistic
is the consideration of statistical noise (variance), but
the challenge behind it is how to estimate the variance from
a small number of samples \citep{chen10}. Since variance is calculated around the
mean which is also estimated, one idea for robust variance
estimation is to iteratively remove outliers then calculate
mean and variance \citep{igor}. The drawback of this approach
is that the number of samples used is further reduced.
Artificially increasing the sample size by resampling (Bootstrapping)
has been considered \citep{braga}. Yet another approach is
to use non-parametric tests in place of the $t$-test
(e.g. Mann-Whitney-Wilcoxon test), so that the variance
estimation is not required.

The line of thoughts we pursue for a robust variance estimation 
is motivated by the typical ``large $p$ small $n$" situation 
for a microarray experiment \citep{liyang}. Though the sample 
size $n$ could be small, the number of genes $p$ is nevertheless 
large, and that large number of genes make it possible for 
a reliable estimation of common variance cross all genes 
\citep{pan,wright,jain,cui05}, at least for the control group.

One main worry about variance estimation is that its value can
be low due to the low expression level. To avoid the estimated
variance being too low, we may add a constant ``penalty"
term $s_0$ to the sample-estimated standard deviation \citep{sam}
(under a not-so-informative name ``SAM" for significance analysis
of microarrays): 
\begin{equation}
\label{eq-sam}
 t_{sam} = \frac{ \langle \log_{10} E_1 \rangle - \langle \log_{10} E_0 \rangle}
{ \sqrt{\frac{s_1^2}{n_1} + \frac{s_0^2}{n_0} } +s_0}.
\end{equation}
The penalty is also called ``regularization", reflecting the
prior belief (in the Bayesian framework) that variance estimation
across different genes should exhibit certain smooth behavior
\citep{baldi,hastie}.

A popular software package called SAM (Significance Analysis of Microarrays)
\citep{sam-docu}
({\sl http://www-stat.stanford.edu/\~{}tibs/SAM/}, version 4.0, July 2010) 
is based on Eq.(\ref{eq-sam}).
Another {\sl R} implementation of the same idea, {\sl siggenes} \citep{siggenes}, 
is available at \\ 
{\sl http://www.bioconductor.org/packages/release/bioc/html/siggenes.html}.
In SAM \citep{sam-docu},  the $s_0$ value is chosen to minimize the 
variability of $t_{sam}$ with respect to the gene-specific standard error
term of $\sqrt{s_1^2/n_1+s_0^2/n_0}$. In \citep{efron}, $s_0$ is
set at the 90\% percentile of standard errors of all genes.
In practice, any small value of $s_0$ can stabilize the variance
estimation.

A Bayesian derivation of the extra term in variance estimation
is derived in \citep{baldi}. In this framework, mean, variance
of a normal distribution (of $\log_{10}(x)=x'$) has a prior distribution,
as well as a posterior distribution after data are observed. For convenience,
the inverse Gamma distribution for the variance parameter and the
normal distribution for the mean parameter is chosen to ensure  
both prior and posterior distribution to have the same functional form. 
It can be shown that (the mean of) posterior variance is a weighted sum
of prior variance ($\sigma_0^2$) and the sample-estimated 
of variance $s^2$ \citep{baldi}:
\begin{equation}
\label{max_post_var}
E[\sigma_{posterior}^2] = w s^2 + (1-w) \sigma_0^2  
\end{equation}
where weight $w$ ($n$ is the sample size, $\nu_0$ is the prior degree of freedom
for the inverse Gammar distribution):
\begin{equation}
w= \frac{n-1}{\nu_0 +n-2}
\end{equation}
tend to close to 1 for larger sample size.

The moderated or regularized variance $\sigma_{posterior}^2$
in Eq.(\ref{max_post_var}) has the effect of drawing gene-specific
variance towards the middle, since its change from the 
sample estimated variance:
\begin{equation}
\sigma_{posterior}^2- s^2 = w s^2 + (1-w) \sigma_0^2 - s^2
= -(1-w) (s^2- \sigma_0^2),
\end{equation}
is negative when $s^2 > \sigma_0^2$ and positive when
$s^2  < \sigma_0^2$.
Note that in Eq.(\ref{max_post_var}), it is the variance 
that is additive, whereas it is standard error that is 
additive in the denominator of Eq.(\ref{eq-sam}). 
However, the idea of moderation/regularization by
adding an extra positive and constant term to
the sample-estimated one is the same.

In fact, there is a second extra term in variance estimation
if the sample-estimated mean is not a good estimate of the
true mean \citep{baldi}. For this reason, it is reasonable
to consider removing outliers to make sure the mean is
estimated robustly \citep{igor}.

\section{Regularized $t$-statistic as a joint filtering criterion}
\indent

What is the relationship between robust variance estimation
or regularization discussed in the last section
and the volcano plot? If FC can be considered to be the special case
when variances of all genes are equal, $t$-statistic of course 
contains gene-specific variance, then $t_{sam}$ in Eq.(\ref{eq-sam})
is somewhere in-between \citep{zhang}. Rewrite 
$|\langle \log_{10} E_1 \rangle - \langle \log_{10} E_0 \rangle|$
as $\delta$ (log-fold-change), $\sqrt{ s_1^2/n_1 + s_0^2/n_0}$ as $s$
(standard error), the regularized $t$-statistic in Eq.(\ref{eq-sam}) can be
split into two terms \citep{zhang}:
\begin{equation}
\label{eq-sam-weighted-sum}
t_{sam}= \frac{\delta}{s+s_0}
= \frac{1}{2(s+s_0)} \cdot \delta + \frac{s}{2(s+s_0)} \cdot \frac{\delta}{s}
\end{equation}
In other words, $t_{sam}$ is a weighted sum of $\log_{10}(FC')$ and $t$-statistic,
$t_{sam} = a \delta  + b (\delta/s)$,
where $a= 0.5/(s+s_0)$, $b =0.5s/(s+s_0)$.

\begin{figure}[th]
 \begin{center}
 \begin{turn}{-90}
 \epsfig{file=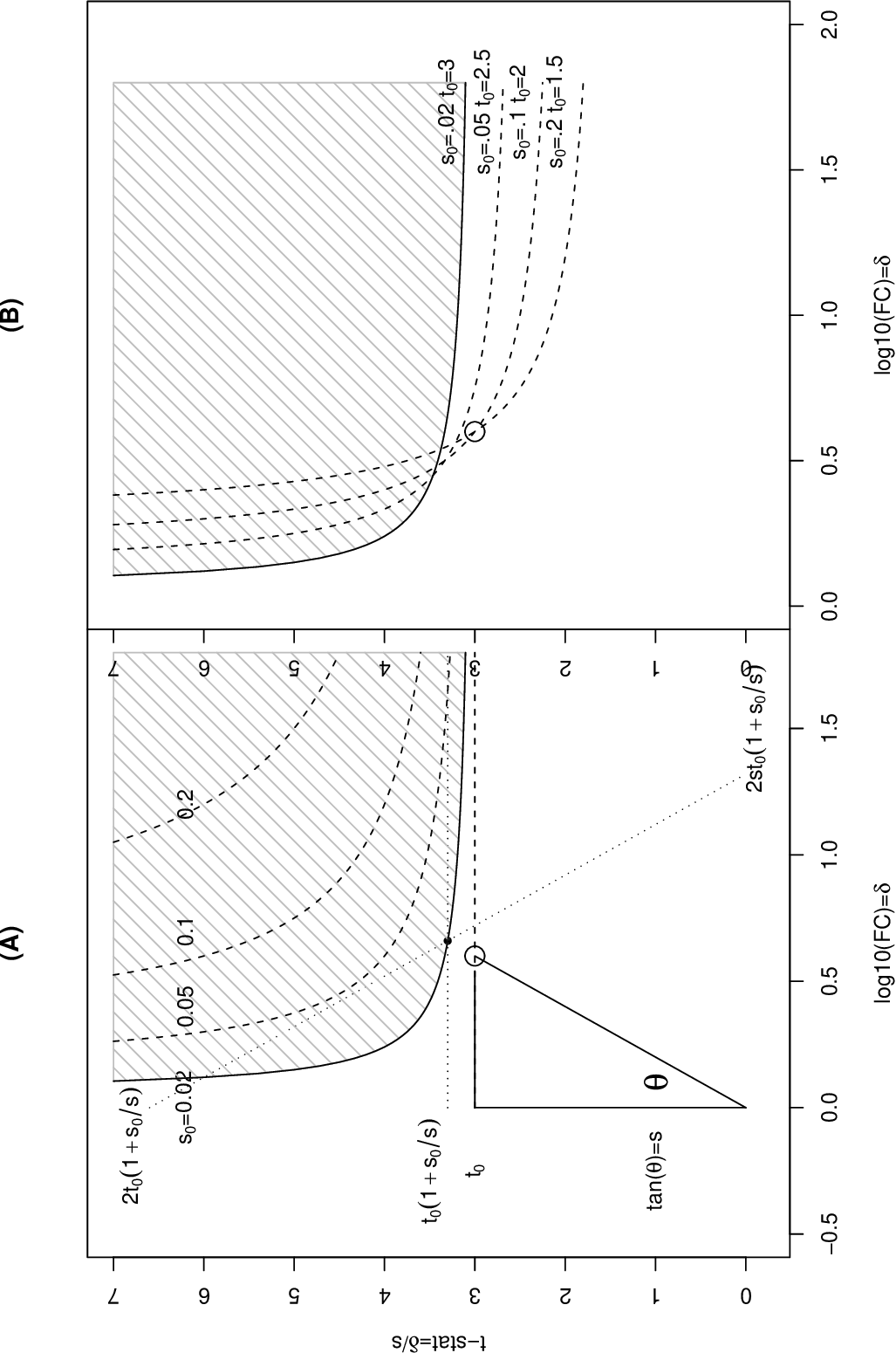, height=11cm}
 \end{turn}
 \end{center}
\caption{
\label{fig7}
Illustration of the regularized $t$-statistic ($t_{sam}$) 
in volcano plot. For a gene with $x=\delta=\log_{10}(FC')=0.6$
and $y=\delta/s=t=3$ (or $s=0.2$, $\theta=5.655^o$),
the forced linear line (under conflicting limits) for
$s_0=0.02$ is shown. Also shown are the discriminant lines
$t_{sam} \ge t_0=3$ at $s_0=0.02, 0.05, 0.1, 0.2$.
(B) Decreasing $t_0$ when $s_0$ is increased:
$s_0=0.02, t_0=3$, $s_0=0.05, t_0=2.5$,
$s_0=0.1, t_0=2$, and $s_0=0.2, t_0=1.5$,
}
\end{figure}

Eq.(\ref{eq-sam-weighted-sum}) might suggest that $t_{sam}$ is a linear
combination of $\log_{10}(FC')$ and $t$, and the gene filtering
criterion $t_{sam} \ge t_0$ discriminant line is a straight line
in the volcano plot. However, this geometric interpretation is
incorrect. The first hint comes from the fact that the split
of $t_{sam}$ into two terms in Eq.(\ref{eq-sam-weighted-sum}) can
also be carried out for $t$ itself: $t= (1/2s) \delta + (1/2) t$.
This is apparently paradoxical as $t \ge t_0$ without regularization
should be the plane above the line of $y=t_0$, without a
contribution from the $x$-axis.  The second hint is from the 
observation that the coefficients of ``linear function" ($a$ and $b$) 
are not constants, but function of the variables themselves. 

The third hint can be seen if you want to draw an actual discriminant 
straight line: the $y$-intercept is obtained in the limit of
$\delta \rightarrow 0$, $s \rightarrow 0$, but $\delta/s >0$.
Since $x/y= \delta/(\delta/s)=s$, the standard error $s$ has a
simple geometric meaning as $\tan(\theta)$ where $\theta$ is the
angle between the $y$-axis and the line linking the point and the origin.
The above $s \rightarrow 0$ limit corresponds to the point to
move closer to the $y$-axis. Similarly, in order to obtain the
$x$-intercept, the limits to be taken are $\delta/s \rightarrow 0$,
$\delta >0$, and $s \rightarrow \infty$. This is the limit for
the point to move away from the $y$-axis to infinity. Interestingly,
under these two conflicting limits, both $y$- and $x$-intercept
can be obtained: $y$-intercept equal to $2t_0 (1+s_0/s)$,
$x$-intercept equal to $2st_0 (1+s_0/s)$ (see Fig.\ref{fig7}(A)).

\begin{figure}[th]
 \begin{center}
 \begin{turn}{-90}
 \epsfig{file=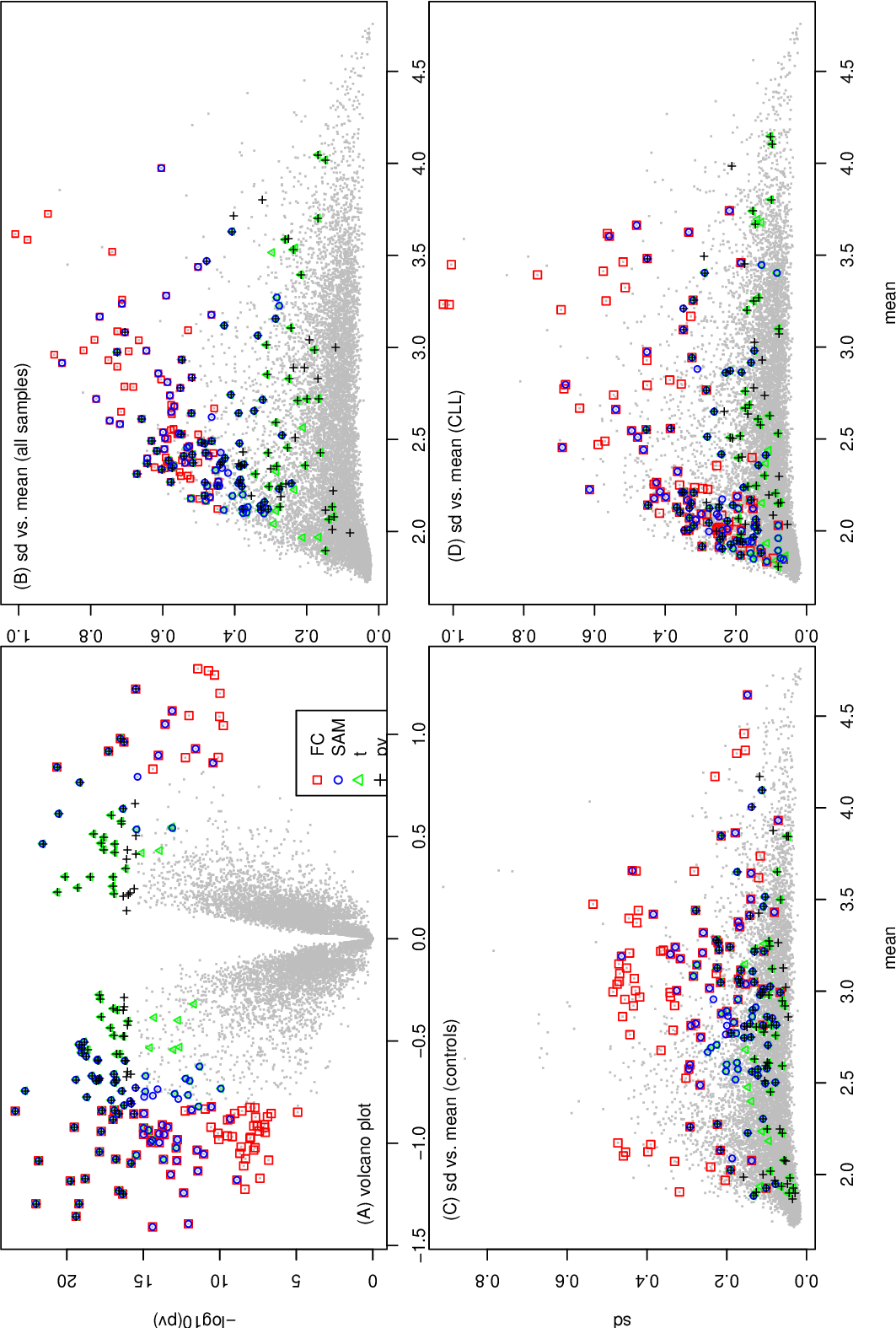, height=17cm}
 \end{turn}
 \end{center}
\caption{
\label{fig8}
Green, red, green, black dots are the top 100 probes/genes selected
by $t_{sam}$, $FC'$, $t$-statistic, and $p$-value of $t$-test.
(A) on volcano plot, $x$: $\log_{10}(FC')$, $y$: $-\log_{10}(p$-value).
(B) $x$: mean of all samples, $y$: standard deviation of all samples.
(C) $x$: mean of control samples, $y$: standard deviation of control samples.
(D) $x$: mean of diseased samples, $y$: standard deviation of diseased samples.
}
\end{figure}

The correct decomposition of $t_{sam}$ splits it into $t$ and $s$:
\begin{eqnarray}
\label{eq-curve}
t_{sam} = \frac{\delta}{s+s_0}= \frac{\delta}{s} (1+ \frac{s_0}{s})^{-1} &\ge& t_0\nonumber \\
\mbox{or,} \hspace{0.2in}
t &\ge & t_0 (1+ \frac{s_0}{s})
\end{eqnarray}
In other words, the discriminant line is a curve which moves up
for smaller $s$'s (smaller angles, smaller FC's). A large 
$t$-statistic but a small FC (the $t$-filtering only area in
Fig.\ref{fig4}) is more difficult to pass 
the filtering in Eq.(\ref{eq-curve}). And a large FC without
a minimum $t$-statistic ($t_0$) (the FC-filtering only area
in Fig.\ref{fig4}) would not pass the filtering 
no matter what. These are very different conclusions when
compared to the linear discriminant line illustrated in
Fig.\ref{fig7}(A) where a large FC but a small $t$ may still
be selected.

Fig.\ref{fig7} also illustrates the effect of $s_0$.
Besides the discriminant lines at $s_0=0.02$, three
more lines are shown at $s_0=0.05, 0.1$, and 0.2,
or 20\%, 50\%, and 100\% of $s$. 
We may increase $s_0$ while decrease $t_0$ at
the same time so that these lines are similar, as shown in
Fig.\ref{fig7}(B). Under the condition that the same 
number of top-ranking genes are selected, the exact 
value of $s_0$ is less important than the fact that this 
term is added ($s_0 > 0$), though Fig.\ref{fig7}(B) does
show that with a larger $s_0$ value, more genes with
less significance but larger FCs are selected.

Fig.\ref{fig8}(A) compares the top 100 genes selected by
SAM (regularized $t$) (blue) with those selected by FC (red),
$t$-test $p$-value (black), and $t$-statistic itself (green).
Although there are certain overlaps among different selection 
criteria, SAM is able to pick up genes that are not selected
by either FC or $t$-test $p$-value alone.
To address the question on whether $t$-test criterion
tends to select genes with low variance and low expression level.
Fig.\ref{fig8}(B)(C)(D) show the standard deviation ($y$-axis)
vs. mean ($x$-axis) for all samples, control samples only,
and diseased (CLL) samples only. Indeed, FC-based criterion
tend to select genes with high variances, $t$-test based
criterion selects relatively low variance genes, and SAM achieves 
a balance between the two criteria, selecting genes with 
intermediate variance values. On the other hand, there is
no strong evidence that any selection criterion tends to select low 
expression level genes.

\section{Relevant Bioconductor programs}
\indent

There are many commercial microarray data analysis programs
that include volcano plots. There are also many general graphic
packages that intend to handle large number of points, such
as {\sl ggplot2} \citep{ggplot2}. To make our discussion managable,
we limit our summary to Bioconductor programs.
Bioconductor site (version 2.10) is a major repository of microarray
analysis softwares written in {\sl R} ({\sl http://www.r-project.org/})
\citep{biocond04,biocond}.
Table \ref{table1} lists packages that are relevant to the
discussions in this review, roughly grouped into three types:
\begin{itemize}
\item 
{\sl volcano plots:} These are straightforward implementations
of the scatter plot, with $x$-axis usually the log-fold-change
and $y$-axis any other measure of differential expression.
We list packages not only for analyzing fluorescence-light 
intensity-based mRNA expression data (Affymetrix, Illumina, etc.),
but also for mRNA expression level based on count data (RNA-seq), 
and protein expression levels, etc.

\item
{\sl alternative measures of differential expression:} The fold-change
and $t$-statistic (or $-\log_{10}(p$-value), or 
the regularized $t$-statistic (SAM), are not the only measures of 
differential expression. One large group of alternative 
measures is the Bayesian calculation of the posterior 
probability that a gene belong to the differential
expression subset (e.g. empirical Bayes analysis of microarray (EBAM) ).
The packages {\sl DEDS} \citep{deds} and {\sl GeneSelector} \citep{boule}, 
in particular, include large number of these measures (F-statistic, B-statistic, 
moderated-F, moderated-t, shrinkage-t, etc.).

\item
{\sl improvement on error/variance estimation:} Robust
and reliable variance estimations are at the heart of the
dichotomy choices between $t$ and FC. Some functions in 
Bioconductor packages directly address this issue, and
are listed in Table \ref{table1}. For example, using 
variation among replicated samples, using variance 
between similar probesets, pooling errors, removing outliers, 
etc.

\end{itemize}

\begin{table}
\begin{center}
\begin{tabular}{l|l|l}
\hline
package & functions & comments\\
\hline
a4\citep{a4} & volcanoPlot, topTable, limmaTwoLevels& 
 A4 for ``automatic Affymetrix array analysis"\\
ABarray & doPlotFCT, doLPE & 
 AB for ``Applied Biosystems" \\
 & & FCT for ``fold-change and t-statistic" \\
 & & DEDS for ``differential expression via distance synthesis"\\
ArrayTools & selectSigGene & double filtering criterion \\
baySeq\citep{bayseq} & plotPosteriors & Bayesian, RNA-seq\\
cummeRbund & csVolcano & \\
DEDS\citep{deds} & deds.stat/deds.stat.linkC & include:
  t, F, FC, SAM, modt, modF, B \\
 & & DEDS for ``differential expression via distance synthesis"\\
 & & SAM for ``significance analysis of microarrays"\\
DEGseq\citep{degseq} & samWrapper & RNA-seq \\
DESeq\citep{deseq} & estimateDispersion, nbinomTest & RNA-seq \\
diffGeneAnalysis\citep{dga} & biasAdjust & \\
GeneSelector\citep{boule}  & RankingBaldiLong/Ebam/FC/FoxDimmic, &
 EBAM for ``empirical Bayes analysis of microarrays" \\
& FoxDimmic/Limma/Permutation/Sam, & the function names preceded by ``Ranking" \\
& ShrinkageT/SoftthresholdT/Tstat & \\ 
& WelchT/WilcEbam/Wilcoxon, & \\
limma\citep{limma} & lmFit, eBayes, volcanoplot, topTable & Bayesian\\
maanova\citep{maanova} & volcano & \\
maDB & drawVolcanoPlot & \\
nudge\citep{nudge} & nudge1 & Bayesian \\
oneChannelGUI & dfMAPlot & draw from limma's topTable\\
pickgene\citep{pickgene} & pickgene & \\
plgem\citep{plgem} & plgem.obsStn, plgem.deg & 
PLGEM for ``power law global error model" \\
 & & STN for ``signal-to-noise (ratio)" \\
 & & DEG for ``differentially expressed genes"\\
PLPE\citep{plpe} & lpe.paired & protein level. PLPE for ``paired local pooled error"\\
plw\citep{plw} & plw, topRankSummary & PLW for ``probe-level locally-moderated weighted (t-test)"\\
puma\citep{puma} & pumaDE, calculateLimma, topGenes & 
PU for ``propagating uncertainty"\\
RankProd\citep{rankprod} & RP, RPadvance, topGene & RP for ``rank product"\\ 
SAGx\citep{broberg} & samrocN &  \\
samr & & SAM. not distributed through bioconductor \\
siggenes\citep{siggenes} & sam, d.stat, ebam & SAM, Bayesian \\
XDE\citep{xde} & xde, calculateBayesianEffectSize & Bayesian \\
xps$^{17}$ & plotVolcano& XPS for ``eXpression Profiling System"\\
\hline
\end{tabular}
\end{center}
\caption{\label{table1}
}
\end{table}

\section{Stratified volcano plots by external information}
\indent

Volcano plot is a 2-dimensional graphic tool, with potentially 
interesting genes scattered outward away from the origin. We
can make volcano plots even more useful by coloring points with
external information. If that external piece of information is
relevant to differential expression, we can easily recognize
the fact by a visual impression of the plot. This coloring of
a volcano plot can be called ``stratified volcano plot". One
example is to label all probes/genes that belong to a particular
pathway, cellular component, function, or process coded in
gene ontology (GO) categories \citep{go}.

\begin{figure}[th]
 \begin{center}
 \begin{turn}{-90}
 \epsfig{file=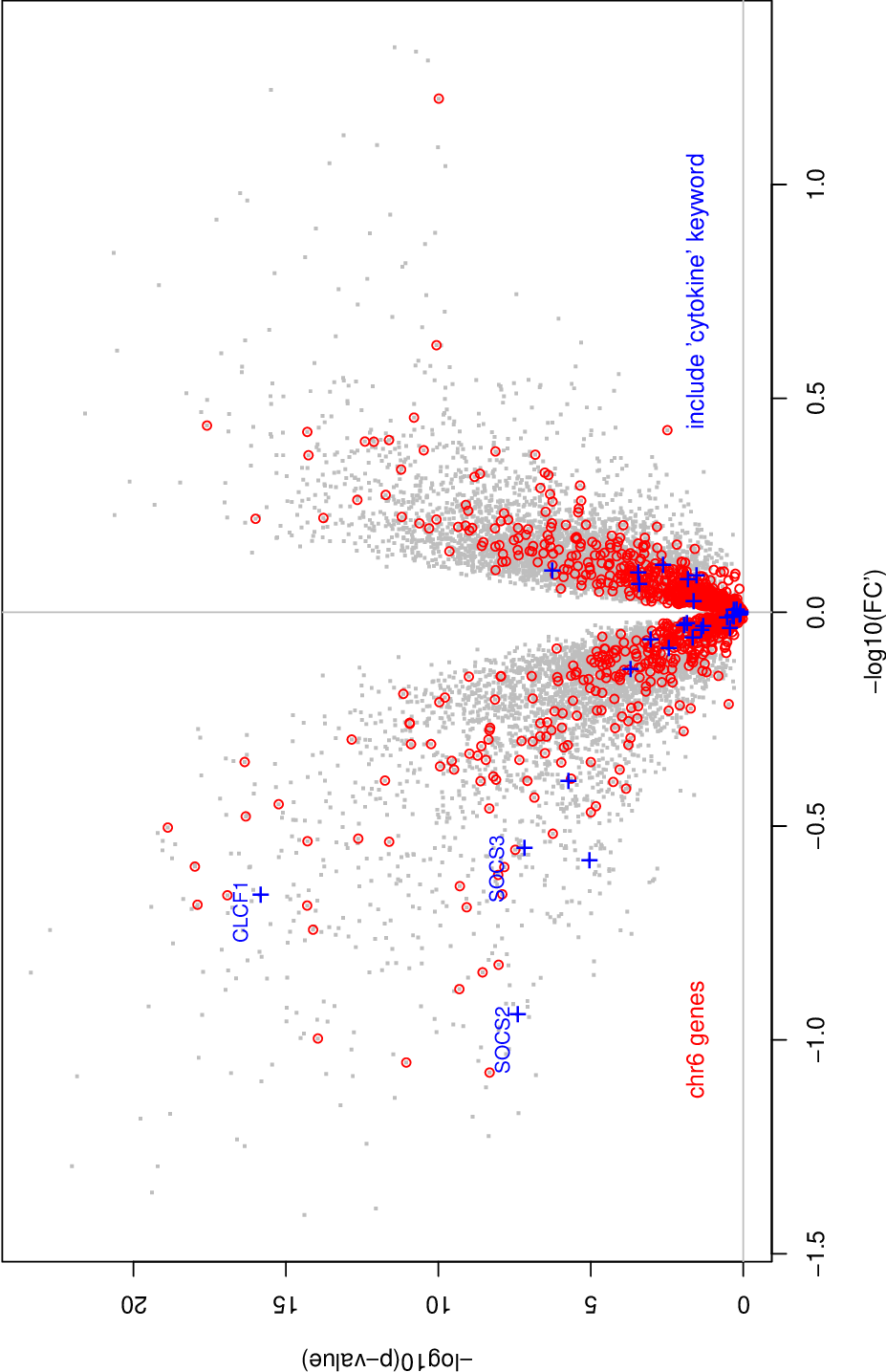, height=13cm}
 \end{turn}
 \end{center}
\caption{
\label{fig9}
Stratified volcano plot: probes/genes on chromosome 6 are marked
by red, and those with ``cytokine" in gene annotation is marked
by blue.
}
\end{figure}

Fig.\ref{fig9} illustrates a stratified volcano plot by marking
1614 probes/genes that are located on chromosome 6 (red),
and 31 probes/genes whose annotation contains the word ``cytokine".
From the stratified volcano plot, we can easily identify
interesting candidate genes involving cytokines such
as CLCF1 (cardiotrophin-like cytokine factor 1, $p$-value= 1.4$\times 10^{-16}$,
FC'=0.22), SOCS2 (suppressor of cytokine signaling 2, 
FC'= 0.11, $p$-value= 3.8$\times 10^{-8}$), SOCS3 (suppressor 
of cytokine signaling 3, FC'=0.28, $p$-value= 6.2 $\times 10^{-8}$), etc.
The visual impression immediately shows the top-ranking
cytokine-linked genes are all down-regulated instead of
up-regulated.

\section{Discussion and conclusions}
\indent

Finding effect measure of differential expression remains an
active research topic \citep{pivalue}. However,
the idea of regularization (adding a small positive term
to the observed standard error to standardize the differential
expression signal) has already successfully combined the two most
well known quantities, log-fold-change and $t$-statistic,
in gene filtering. And volcano plot is a natural visual tool
to illustrate this procedure.

Simultaneous displaying of noise-level-standardized signal 
and unstandardized one can also be useful beyond the field of microarray.
In genetic association studies, the association signal of
a single-nucleotide polymorphism (SNP) is usually measured by
two quantities. One is the odds-ratio (OR) of the 2-by-2 count table
with disease status as row and two alleles as column. OR is not
standardized by the noise level or sample size, though the 95\%
confidence interval of OR does become narrower for larger sample
sizes thus lower level of chance events \citep{li-bib}. On the
other hand, the chi-square statistic or the $p$-value of the
chi-square ($\chi^2$) test strongly dependent on sample size,
thus chance event probability. In fact, the chi-square statistics
is proportional to the total number of samples for a SNP that
contains association signals.

Besides using OR in $x$-axis (in log scale), another choice is
to use the allele frequency difference in case and control group.
Denote the four counts in the 2-by-2 table (row for case control status,
columns for presence of absence of a particular allele/genotype) 
in case-control association analysis are $a,b,c,d$, $\log_{10}$OR is 
$\log_{10}(ad)-\log_{10}(bc)$, whereas allele frequency difference 
is $a/(a+b)- c/(c+d)= (ad-bc)(a+b)^{-1}(c+d)^{-1}$. In other
words, the difference between the two choices is whether 
$ad$ and $bc$ are compared in the logarithmic or regular scale.

It is rare for volcano plots being applied to genetic association
studies (some examples can be found in \citep{sirota,miclaus}). 
We believe that many extensions
and applications of volcano plots in microarray analysis
can be equally useful in genetic association analysis.
For example, the joint filtering criterion, the stratified
volcano plot coloring external pieces of information,
and uncovering of systematic patterns when points are colored 
by other information. We have found that the location 
of a SNP on the volcano plot is intrinsically related 
to its minor allele frequency. This will provide further 
insight on how one should balance the chi-square test result
and odds-ratio in selecting genetically associated genes.

In conclusion, volcano plot, together with heatmaps
\citep{eisen}, MA plots \citep{ma}, 
and cluster/PCA plots \citep{pca,leek10},  is among the most useful and
most frequently used visual tools in microarray analysis,
Volcano plots display both noise-level-standardized
and unstandardized signal concerning differential expression
of mRNA levels. Regularized test statistic and joint filtering 
have an intuitive geometric interpretation in volcano plot, 
and its advantage over double filter criterion of genes 
can be easily understood.  As a scattering plot, volcano 
plot can incorporate other external information, such as 
gene annotation, to aid the hypothesis generating process 
concerning a disease or phenotype.

\vspace{-0.2in}

\section*{Acknowledgements}

I would like to thank Frank Batliwalla, Percio Gulko, Max Brenner,
Peter Gregersen for asking about differences between genes selected
by fold-change and those by $t$-test, which provided the initial 
incentive for writing this review, Joy Yan, Sophia Yancopoulos 
for discussion on volcano plots and allowing me to use the CLL data, 
Yaning Yang, Xing Hua for collaboration on software development,
and Michaela Oswald for reading a draft of the paper.


\normalsize

\begin{thebibliography}{99}

\bibitem[Schena et al.,(1998)]{schena}
M. Schena, R.A. Heller, T.P. Theriault,  K. Konrad, E. Lachenmerier, R.W. Davis
``Microarrays: biotechnology's discovery platform
for functional genomics",
{\sl Trends in Biotech.} {\bf 16},301-306 (1998).

\bibitem[Young,(2000)]{young}
R.A. Young,
``Biomedical discovery with DNA arrays",
{\sl Cell} {\bf 102}, 9-15 (2000).

\bibitem[Butte,(2002)]{butte}
A. Butte ,
``The use and analysis of microarray data",
{\sl Nat. Rev. Drug Discovery} {\bf 1}, 951-960 (2002).

\bibitem[Slonim,(2002)]{slonim}
D.K. Slonim ,
``From patterns to pathways: gene expression data
analysis comes of age",
{\sl Nat. Genet.} {\bf 32(suppl)}, 502-508 (2002).

\bibitem[Stoughton,(2005)]{stoughton}
R.B. Stoughton ,
``Application of DNA microarrays in biology",
{\sl Ann. Rev. Biochem.} {\bf 74}, 53-82 (2005).

\bibitem[Trevino et al.,(2007)]{trevino}
V. Trevino, F. Falciani, H.A. Barrera-Saldana,
``DNA Microarrays: a powerful genomic tool for biomedical and clinical research",
{\sl Mol. Med.} {\bf 13}, 527-541 (2007).

\bibitem[Trachtenberg et al.,(2012)]{trachtenberg}
A.J. Trachtenberg, J.H. Robert, A.E. Abdalla, A. Fraser, S.Y. He, 
J.N. Lacy, C. Rivas-Morello, A. Truong, G. Hardiman, L. Ohno-Machado, F. Liu, E. Hovig, W.P. Kuo,
``A primer on the current state of microarray technologies",
{\sl Methods in Mol. Biol.} {\bf 802}, 3-17 (2012).

\bibitem[Etienne et al.,(2004)]{etienne}
W. Etienne, M.H. Meyer, J. Peppers,  R.A. Meyer Jr.,
``Comparison of mRNA gene expression by RT-PCA and
DNA microarray"
{\sl BioTechniques} {\bf 36}, 618-626 (2004).

\bibitem[Dallas et al.,(2005)]{dallas}
P.B. Dallas, N.G. Gottardo, M.J. Firth, A.H. Beesley, K. Hoffmann, 
P.A. Terry, J.R. Freitas, J.M. Boag, A.J. Cummings,  U.R. Kees,
``Gene expression levels assessed by oligonucleotide microarray analysis 
and quantitative real-time RT-PCR -- how well do they correlate?",
{\sl BMC Genomics} {\bf 6}, 59 (2005).

\bibitem[Morey et al.,(2006)]{morey}
J.S. Morey, J.C. Ryan, F.M. van Dolah,
``Microarray validation: factors influencing correlation between oligonucleotide 
microarrays and real-time PCR", 
{\sl Biological Procedures Online} {\bf 8}, 175-193 (2006).

\bibitem[Park et al.,(2004)]{park}
P.J. Park, Y.A. Cao, S.Y. Lee, J.W. Kim, M.S. Chang, R. Hart, S. Choi,
``Current issues for DNA microarrays: platform comparison, double 
linear amplification, and universal RNA reference",
{\sl J. Biotech.} {\bf 112}, 225-245 (2004). 

\bibitem[Larkin et al.,(2005)]{larkin}
J.E. Larkin, B.C. Frank, H. Gavras, R. Sultana, J. Quackenbush,
``Independence and reproducibility across microarray platforms",
{\sl Nat. Methods} {\bf 2}, 337-344 (2005).

\bibitem[Irizarry et al.,(2005)]{irizarry}
R.A. Irizarry,  D. Warren, F. Spencer, et al.,
``Multiple-laboratory comparison of microarray platforms",
{\sl Nat. Methods} {\bf 2}, 345-350 (2005).

\bibitem[Draghici et al.,(2006)]{draghici}
S. Draghici, P. Khatri, A.C. Eklund, Z. Szallasi,
``Reliability and reproducibility issues in DNA microarray measurements",
{\sl Trends in Genet.} {\bf 22}, 101-109 (2006).

\bibitem[Kuo et al.,(2006)]{kuo}
W.P. Kuo,  F. Liu, J. Trimarchi,  et al.,
``A sequence-oriented comparison of gene expression measurements across 
different hybridization-based technologies",
{\sl Nat. Biotech.}, {\bf 24}, 832-840 (2006).

\bibitem[Patterson et al.,(2006)]{patterson}
T.A. Patterson,  E.K. Lobenhofer, S.B. Fulmer-Smentek, et al.,
``Performance comparison of one-color and two-color platforms within the 
Microarray Quality Control (MAQC) project",
{\sl Nat. Biotech.} {\bf 24}, 1140-1150 (2006).

\bibitem[Chen et al.,(2007)]{chen07}
J.J. Chen, H.M. Hsueh, R.R. Delongchamp, C.J. Lin, C.A. Tsai,
``Reproducibility of microarray data: a further analysis of microarray 
quality control (MAQC) data",
{\sl BMC Bioinf.} {\bf  8}, 412 (2007).

\bibitem[Wen et al.,(2011)]{wen}
Z. Wen, Z. Su, J. Liu, B. Ning, L. Guo, W. Tong, L. Shi,
``The MicroArray Quality Control (MAQC) project and cross-platform
analysis of microarray data",
in {\sl Handbook of Statistical Bioinformatics}, 
eds. H Horng-Shing, B Sch\"{o}lkopf, H Zhao, {\bf 2011}, pp.171-192 (2011).

\bibitem[Li et al.,(2012)]{zli12}
Z. Li, J.C. Kwekel, T. Chen ,
``Functional comparison of microarray data across multiple platforms using 
the method of percentage of overlapping functions",
{\sl Methods in Mol. Biol.} {\bf 802}, 123-139 (2012).

\bibitem[Allen et al.,(2012)]{allen}
J.D. Allen, S. Wang, M. Chen, L. Girard, J.D. Minna, Y. Xie, G. Xiao,
``Probe mapping across multiple microarray platforms",
{\sl Brief. Bioinf.}, in press (2012).

\bibitem[Churchill, (2002)]{churchill}
G.A. Churchill,
``Fundamentals of experimental design for cDNA microarrays",
{\sl Nat. Genet.} {\bf 32}, 490-495 (2002).

\bibitem[Li and Rabinovic,(2007)]{cli}
C. Li, A. Rabinovic,
``Adjusting batch effects in microarray expression data using
empirical Bayes methods",
{\sl Biostat.} {\bf 8}, 118-127 (2007).

\bibitem[Baggerly et al.,(2008)]{baggerly}
K.A. Baggerly, K.R. Coombes, E.S. Neeley,
``Run batch effects potentially compromise the usefulness of genomic 
signatures for ovarian cancer"
{\sl J. Clinical Oncology} {\bf 26}, 1186-1187 (2008).

\bibitem[Kitchen et al.,(2010)]{kitchen}
R.R. Kitchen, V.S. Sabine, A.H. Sims, 
E.J. Macaskill, L. Renshaw, J.S. Thomas, J.I. van Hemert, J.M. Dixon, J.M.S. Bartlett,
``Correcting for intra-experiment variation in Illumina BeadChip data 
is necessary to generate robust gene-expression profiles",
{\sl BMC Genomics} {\bf 11}, 134 (2010).

\bibitem[Chen et al.,(2011)]{cchen}
C. Chen, K. Grennan, J. Badner, D. Zhang, E. Gershon, L. Jin, C. Liu,
``Removing batch effects in analysis of expression microarray data: 
an evaluation of six batch adjustment methods",
{\sl PLoS ONE} {\bf 6}, e17238 (2011).

\bibitem[Gagnon-Bartsch and Speed,(2012)]{gagnon}
J.A. Gagnon-Bartsch, T.P. Speed,
``Using control genes to correct for unwanted variation in microarray data",
{\sl Biostat.} {\bf 13}, 539-552 (2012).

\bibitem[Ioannidis,(2005)]{ioannidis}
J.P.A. Ioannidis,
``Microarrays and molecular research: noise discovery?",
{\sl Lancet} {\bf 365}, 454-455 (2005).

\bibitem[Raser ad O'Shea,(2005)]{raser}
J.M. Raser, E.K. O'Shea,
``Noise in gene expression: origins, consequences, and control",
{\sl Science} {\bf 309}, 2010-2013 (2005).

\bibitem[Ein-Dor et al.,(2006)]{eindor}
L. Ein-Dor, O. Zuk, E. Domany,
``Thousands of samples are needed to generate a robust gene list for predicting outcome in cancer",
{\sl Proc. Nat. Acad. Sci. } {\bf 103}, 5923-5928 (2006).

\bibitem[Maheshri and O'Shea,(2007)]{maheshri}
N. Maheshri, E.K. O'Shea,
``Living with noisy genes: how cells function reliably with inherent variability in gene expression",
{\sl Annu. Rev. Biophy. Biomol. Struct.} {\bf 36}, 413-434 (2007).

\bibitem[Zeisel et al.,(2010)]{zeisel}
A. Zeisel, A. Amir, WJ K\"{o}stler, E. Domany,
``Intensity dependent estimation of noise in microarrays improves 
detection of differentially expressed genes",
{\sl BMC Bioinf.} {\bf 11}, 400 (2010).

\bibitem[Thomas et al.,(2010)]{thomas}
R. Thomas, L. de la Torre, X. Chang, S. Mehrotra,
``Validation and characterization of DNA microarray gene expression data 
distribution and associated moments",
{\sl BMC Bioinf.} {\bf 11}, 576 (2010).

\bibitem[Kitchen et al.,(2011)]{kitchen11}
RR Kitchen, VS Sabine, AA Simen, J.M. Dixon, J.M.S. Bartlett, A.H. Sims,
``Relative impact of key sources of systematic noise in Affymetrix and Illumina 
gene-expression microarray experiments",
{\sl BMC Genomics} {\bf 12}, 589 (2011).

\bibitem[Posekany et al.,(2011)]{posekany}
A. Posekany, K. Felsenstein, P. Sykacek,
``Biological assessment of robust noise models in microarray data analysis",
{\sl Bioinf.} {\bf 27}, 807-814 (2011).

\bibitem[Tang and Yan,(2012)]{tang}
V.T.Y. Tang, H. Yan,
``Noise reduction in microarray gene expression data based 
on spectral analysis",
{\sl Int. J. Machine Learning and Cybernetics} {\bf 3}, 51-57 (2012).

\bibitem[Sharov et al.,(2004)]{sharov}
V. Sharov, K.Y. Kwong, B. Frank, E. Chen, J. Hasseman, R. Gaspard, 
Y. Yu, I. Yang, J. Quackenbush,
``The limits of log-ratios",
{\sl BMC Biotech.} {\bf 4}, 3 (2004).

\bibitem[Yang and Speed,(2002)]{yang}
Y.H. Yang, T. Speed ,
``Design issues for cDNA microarray experiments",
{\sl Nat. Rev. Genet.} {\bf 3}, 579-588 (2002).

\bibitem[Shi et al.,(2004)]{shi}
L. Shi, W. Tong, F. Goodsaid, F.W. Frueh, H.  Fang, T. Han, 
J.C.  Fuscoe, D.A.  Casciano,
``QA/QC: challenges and pitfalls facing the microarray community and regulatory agencies",
{\sl Expert Rev. Mol. Diagnostics} {\bf 4}, 761-777 (2004).

\bibitem[Shi et al.,(2006)]{shi06}
L. Shi, L.H. Reid, W.D. Jones, et al.
MAQC Consortium (2006),
``The MicroArray Quality Control (MAQC) project shows inter- and intraplatform reproducibility of 
gene expression measurements",
{\sl Nat. Biotech.} {\bf 24}, 1151-1161 (2006).

\bibitem[McCall et al.,(2011)]{mccall}
M.N. McCall, P.N. Murakami, M. Lukk, W. Huber, R.A. Irizarry,
``Assessing affymetrix GeneChip microarray quality",
{\sl BMC Bioinf.} {\bf 12}, 137 (2011).

\bibitem[Brazma et al.,(2001)]{miame}
A. Brazma,  P. Hingamp, J. Quackenbush, et al.,
``Minimum information about a microarray experiment (MIAME) --
toward standards for microarray data",
{\sl Nat. Genet.} {\bf 29}, 365-371 (2001).

\bibitem[Ioannidis et al.,(2009)]{ioannidis09}
J.P.A. Ioannidis, D.B. Allison, C.A. Ball, et al.,
``Repeatability of published microarray gene expression analyses",
{\sl Nature Genet.} {\bf 41}, 149-155 (2009).

\bibitem[Quackenbush,(2002)]{quack}
J. Quackenbush,
``Microarray data normalization and transformation",
{\sl Nat. Genet.} {\bf 32}, 496-501 (2002).

\bibitem[Vandesompele et al.,(2002)]{vande}
J. Vandesompele, K de Preter, F. Pattyn, B. Poppe, N. van Roy, 
A. de Paepe, F. Speleman,
``Accurate normalization of real-time quantitative RT-PCR data by geometric 
averaging of multiple internal control genes",
{\sl Genome Biol.} {\bf 3}, 34 (2002).

\bibitem[Fujita et al.,(2006)]{fujita}
A. Fujita, J.R. Sato, L. de Oliveira Rodrigues, C.E. Ferreira,  M.C. Sogayar,
``Evaluating different methods of microarray data normalization",
{\sl BMC Bioinf.} {\bf 7}, 469 (2006).


\bibitem[Steinhoff and Vingron,(2006)]{steinhoff}
C. Steinhoff, M. Vingron,
``Normalization and quantification of differential expression in gene expression microarrays",
{\sl Brief. Bioinf.} {\bf 7}, 166-177 (2006).

\bibitem[Stafford,(2008)]{stafford}
P. Stafford, ed.
{\sl Methods in Microarray Normalization} 
(CRC Press, 2008).

\bibitem[Autio et al.,(2009)]{astola}
R. Autio, S. Kilpinen, M. Saarela, O. Kallioniemi, S. Hautaniemi, J. Astola,
``Comparison of Affymetrix data normalization methods using 6,926 experiments across five array generations",
{\sl BMC Bioinf.} {\bf 10(suppl 1)}, S24 (2009).

\bibitem[\"{O}nskog et al.,(2011)]{onskog}
J. \"{O}nskog, E. Freyhult, M. Landfors, P. Ryd\'{e}n, T.R. Hvidsten,
``Classification of microarrays; synergistic effects between normalization, 
gene selection and machine learning",
{\sl BMC Bioinf.} {\bf 12}, 390 (2011).

\bibitem[Halpert and Sanga,(2012)]{halpert}
R.L. Halpert, S. Sanga,
``Robust unattended microarray analysis",
Ingenuity Technical Report No. 2011-1 (Ingenuity Systems, Inc. 2012).

\bibitem[Estrada et al.,(2009)]{estrada}
K Estrada, A Abuseiris, FG Grosveld, A.G. Uitterlindern, T.A. Knoch, F. Rivadeneira,
``GRIMP: a web- and grid-based tool for high-speed analysis of large-scale 
genome-wide association using imputed data",
{\sl Bioinf.} {\bf 25}, 2750-2752 (2009).

\bibitem[Schadt et al.,(2010)]{schadt}
E.E. Schadt, M.D. Linderman, J. Sorenson, L. Lee, G.P. Nolan,
``Computational solutions to large-scale data management and analysis",
{\sl Nat. Rev. Genet.} {\bf 11}, 647-657 (2010).

\bibitem[Hastie et al.,(2001)]{hastie}
T. Hastie, R. Tibshirani, J. Friedman,
{\sl The Elements of Statistical Learning}
(Springer, 2001).

\bibitem[Storey and Tibshirani,(2003)]{storey}
J.D. Storey, R. Tibshirani,
``Statistical significance for genomewide studies",
{\sl Proc. Nat. Acad. Sci.} {\bf 100}, 9440-9445 (2003).

\bibitem[Storey,(2003)]{storey-q}
J.D. Storey,
``The positive false discovery rate: a Bayesian interpretation
and the $q$-value",
{\sl Annals of Statistics} {\bf 31}, 2013-2035 (2003).

\bibitem[Reiner et al.,(2003)]{reiner}
A. Reiner, D. Yekutieli, Y. Benjamini,
``Identifying differentially expressed genes using false discovery 
rate controlling procedures",
{\sl Bioinf.} {\bf 19}, 368-375 (2003).

\bibitem[Pawitan et al.,(2005)]{pawitan}
Y. Pawitan, S. Michiels, S. Koscielny,  A. Gusnanto, A. Ploner,
``False discovery rate, sensitivity and sample size for microarray studies",
{\sl Bioinf.} {\bf 21}, 3017-3024 (2005).

\bibitem[Schwartzman and Lin,(2011)]{schw}
A. Schwwartzman, X. Lin,
``The effect of correlation in false discovery rate estimation",
{\sl Biometrika} {\bf 98}, 199-214 (2011).

\bibitem[Xing et al.,(2001)]{xing}
E.P. Xing, M.I. Jordan, R.M. Karp,
``Feature selection for high-dimensional genomic microarray data",
in {\sl Proc. Eighteenth Int. Conf. Machine Learning (ICML 2001)},
eds. CE Brodley, AP Danyluk, pp.  601-608
(Morgan Kauffmann, 2001).

\bibitem[Li and Yang,(2002)]{liyang}
W. Li, Y. Yang,
``How many genes are needed for a discriminant microarray data analysis"
in {\sl  Methods of Microarray Data Analysis }
eds. SM Lin and KF Johnson, pp.137-150 (Kluwer Academic, 2002). 

\bibitem[Ambroise and McLachlan,(2002)]{ambroise}
C. Ambroise, G.J. McLachlan,
``Selection bias in gene extraction on the basis of microarray gene-expression data",
{\sl Proc. Nat. Acad. Sci. } {\bf 99}, 6562-6566 (2002).

\bibitem[Guyon and Elisseeff,(2003)]{guyon}
I. Guyon, A. Elisseeff, 
``An introduction to variable and feature selection",
{\sl J. Machine Learning Res.} {\bf 3}, 1157-1182 (2003).

\bibitem[Peng et al.,(2005)]{ding}
H. Peng, F. Long, C. Ding,
``Feature selection based on mutual information criteria of max-dependency, 
max-relevance, and min-redundancy",
{\sl IEEE Trans. Pattern Analysis and Machine Learning} {\bf 27}, 1226-1238 (2005).

\bibitem[Li,(2006)]{wli06}
W. Li,
``The-more-the-better and the-less-the-better",
{\sl Bioinf.} {\bf 22}, 2187-2188 (2006).

\bibitem[Liao and Chin,(2007)]{liao}
J.G. Liao. K.V. Chin,
``Logistic regression for disease classification using microarray data: 
model selection in a large $p$ and small $n$ case",
{\sl Bioinf.} {\bf 23}, 1945-1951 (2007).

\bibitem[Zhao et al.,(2010)]{zhao}
C. Zhao, M.L. Bittner, R.S. Chapkin, E.R. Dougherty,
``Characterization of the effectiveness of reporting lists of 
small feature sets relative to the accuracy of the prior biological knowledge",
{\sl Cancer Informatics} {\bf 9}, 49-60 (2010).

\bibitem[Golub et al.,(1999)]{golub}
T.R. Golub, D.K. Slonim, P. Tamayo, C. Huard, M. Gaasenbeek, 
J.P. Mesirov, H. Coller, M.L. Loh, J.R. Downing,  M.A. Caligiuri, 
C.D. Bloomfield,  E.S. Lander,
``Molecular classification of cancer: class discovery and class 
prediction by gene expression monitoring",
{\sl Science} {\bf 286}, 531-537 (1999).

\bibitem[Hedenfalk et al.,(2001)]{hedenfalk}
I. Hedenfalk, D. Duggan, Y. Chen, et al.
``Gene-expression profiles in hereditary breast cancer", 
{\sl New England J. Medicine} {\bf 344}, 539-548 (2001).


\bibitem[Dhanasekaran et al.,(2001)]{dhan}
S.M. Dhanasekaran, T.R. Barrette, D. Ghosh, R. Shah, S. Varambally, K. Kurachi,
K.J. Pienta, M.A. Rubin,  A.M. Chinnaiyan,
``Delineation of prognostic biomarkers in prostate cancer",
{\sl Nature} {\bf 412}, 822-826 (2001).

\bibitem[Adib et al.,(2004)]{adib}
T.R. Adib, S. Henderson, C. Perrett, 
D. Hewitt, D. Bourmpoulia, J. Ledermann, C. Boshoff,
``Predicting biomarkers for ovarian cancer using gene-expression microarrays",
{\sl Brit. J. Cancer} {\bf 90}, 686-692 (2004).

\bibitem[Yeatman,(2009)]{yeatman}
T.J. Yeatman,
``Predictive biomarkers: identification and verification",
{\sl J. Clinical Oncology} {\bf 27}, 2743-2744 (2009). 

\bibitem[Pomeroy et al.,(2002)]{pomeroy}
S.L. Pomeroy, P. Tamayo, M. Gaasenbeek, et al.
``Prediction of central nervous system embryonal tumour outcome based on gene expression",
{\sl Nature} {\bf 415}, 436-442 (2002).

\bibitem[van de Vijver et al.,(2002)]{vij}
M.J. van de Vijver, Y.D. He, L.J. van't Veer, et al.,
``A gene-expression signature as a predictor of survival in breast cancer",
{\sl New England J. Medicine} {\bf 347}, 1999-2009 (2002).

\bibitem[Colman et al.,(2010)]{colman}
H. Colman, L. Zhang, E.P. Sulman, et al.
``A multigene predictor of outcome in glioblastoma",
{\sl Neuro-Oncology} {\bf 12}, 49-57 (2010). 

\bibitem[Kim and Paike,(2010)]{kim}
C. Kim, S. Paik,
``Gene-expression-based prognostic assays for breast cancer",
{\sl Nat. Rev. Clinical Oncology} {\bf 7}, 340-347 (2010).

\bibitem[Jain et al.,(2003)]{jain}
N. Jain, J. Thatte, T. Braciale, K. Ley, M. O'Connell, J.K. Lee,
``Local-pooled-error test for identifying differentially
expressed genes with a small number of replicated microarrays",
{\sl Bioinf.} {\bf 19}, 1945-1951 (2003).

\bibitem[Chen et al.,(1997)]{chen97}
Y. Chen, E.R. Dougherty, M.L. Bittner,
``Ratio-based decisions and the quantitative analysis of cDNA microarray images"
{\sl J. Biomedical Optics} {\bf 2}, 364-274 (1997).


\bibitem[Baldi and Long,(2001)]{baldi}
P. Baldi, A.D. Long ,
``A Bayesian framework for the analysis of microarray expression data:
regularized $t$-test and statistical inferences of gene changes",
{\sl Bioinf.} {\bf 17}, 509-519 (2001).

\bibitem[Shi et al.,(2005)]{shi05}
L. Shi, W. Tong, H. Fang, U. Scherf, J. Han, R.K. Puri, F.W. Frueh, 
F.M. Goodsaid, L. Guo, Z. Su, T. Han, J.C. Fuscoe, Z.A. Xu, 
T.A. Patterson, H. Hong, Q. Xie, R.G. Perkins, J.J. Chen, D.A. Casciano,
``Cross-platform comparability of microarray technology: Intra-platform consistency and 
appropriate data analysis procedures are essential", 
{\sl BMC Bioinf.} {\bf 6(suppl 2)}, S12 (2005).

\bibitem[Guo et al.,(2006)]{guo}
L. Guo, E.K. Lobenhofer, C. Wang, R. Shippy, S.C. Harris, 
L. Zhang, N. Mei, T. Chen, D. Herman, F.M. Goodsaid,
P. Hurban, K.L. Phillips, J. Xu, X. Deng, Y.A. Sun, W. Tong, Y.P. Dragan, L. Shi, 
``Rat toxicogenomic study reveals analytical consistency across microarray platforms",
{\sl Nat. Biotech.} {\bf 24}, 1162-1169 (2006).

\bibitem[Zhang et al.,(2008)]{mzhang}
M. Zhang, C.  Yao, Z. Guo, J. Zou, L. Zhang, H. Xiao, D. Wang, D. Yang, X. Gong, 
J. Zhu, Y. Li,  
``Apparently low reproducibility of true differential expression discoveries 
in microarray studies",
{\sl Bioinf.} {\bf 24}, 2057-2063 (2008).

\bibitem[Zhang et al.,(2009)]{mzhang2}
M. Zhang, L. Zhang, J. Zou, C. Yao, H. Xiao, Q. Liu, J. Wang, D. Wang, C. Wang, Z. Guo,
``Evaluating reproducibility of differential expression discoveries in microarray studies by considering correlated molecular changes",
{\sl Bioinf.} {\bf  25}, 1662-1668 (2009).


\bibitem[Boulesteix and Slawski,(2009)]{boule}
A.L. Boulesteix, M. Slawski,
``Stability and aggregation of ranked gene lists",
{\sl Brief. Bioinf.} {\bf 10}, 556-568 (2009).

\bibitem[Witten and Tibshirani,(2007)]{witten07}
D.M. Witten, R. Tibshirani,
``A comparison of fold-change and the t-statistic for microarray
data analysis",
Department of Statistics, Stanford University technical report (2007).

\bibitem[P'Hara and Kotze,(2010)]{ohara}
R.B. O'Hara, D.J. Kotze,
``Do not log-transform count data",
{\sl Math. Ecol. Evol.} {\bf 1}, 118-122 (2010).

\bibitem[Fechner,(1860)]{fechner}
G.T. Fechner, {\sl Elemente der Psychophysik} (Leipzig:
Breitkopf und H\"{a}rtel, 1860).

\bibitem[Auer et al.,(2011)]{auer}
P.L. Auer, S. Srivastava, R.W. Doerge,
``Differential expression -- the next generation and beyond",
{\sl Brief. Bioinf.} {\bf 11}, 57-62 (2011).

\bibitem[Robinson et al.,(2010)]{robinson}
M.D. Robinson, D.J. McCarthy, G.K. Smyth,
``{\tt edgeR}: a Bioconductor package for differential
expression analysis of digital gene expression data",
{\sl Bioinf.} {\bf 26}, 139-140 (2010).

\bibitem[Oshlack et al.,(2010)]{oshlack}
A. Oshlack, M.D. Robinson, M.D. Young,
``From RNA-seq reads to differential expression results",
{\sl Genome Biol.} {\bf 11}, 220 (2010).

\bibitem[Anders and Huber,(2010)]{deseq}
S. Anders, W. Huber,
``Differential expression analysis for sequence count data",
{\sl Genome Biol.} {\bf 11}, R106 (2010).

\bibitem[Hardcastle and Kelly,(2010)]{bayseq}
T.J. Hardcastle, K.A. Kelly,
``baySeq: Empirical Bayesian methods for identifying differential 
expression in sequence count data",
{\sl BMC Bioinf.} {\bf 11}, 422 (2010).

\bibitem[Wang et al.,(2010)]{degseq}
L. Wang, Z. Feng, X. Wang, X. Wang, X. Zhang,
``DEGseq: an R package for identifying differentially expressed genes from RNA-seq data",
{\sl Bioinf.} {\bf 26}, 136-138 (2010).

\bibitem[Cumbie et al.,(2011)]{cumbie}
J.S. Cumbie, J.A. Kimbrel, Y. Di, D.W Schafer, L.J. Wilhelm, 
S.E. Fox, C.M. Sullivan, A.D. Curzon, J.C. Carrington, T.C. Mockler, J.H. Chang
``GENE-Counter: a computational pipeline for the analysis of RNA-seq data for 
gene expression differences",
{\sl PLoS ONE} {\bf 6}, e25279 (2011).

\bibitem[Lee et al.,(2011)]{lee11}
J. Lee, Y. Ji, S. Liang, G. Cai, P. M\"{u}ller,
``On differential gene expression using RNA-Seq data",
{\sl Cancer Inform.} {\bf 10}, 205-215 (2011).

\bibitem[Chen et al.,(2011)]{chen11}
Z. Chen, J. Liu, H.K.T. Ng, S. Nadarajah, H.L. Kaufman, J.Y. Yang, Y. Deng,
``Statistical methods on detecting differentially expressed genes for RNA-seq data",
{\sl BMC Sys. Biol.} {\bf 5(suppl 3)}, S1 (2011).

\bibitem[Tarazona et al.,(2011)]{tarazona}
S. Tarazona, F. Garc\'{i}a-Alcalde, J. Dopazo, A. Ferrer, A. Conesa,
``Differential expression in RNA-seq: a matter of depth",
{\sl Genome Res.} {\bf 21}, 2213-2223 (2011).


\bibitem[Witten,(2011)]{witten11}
D.M. Witten,
``Classification and clustering of sequencing data using a Poisson model",
{\sl Ann. Appl. Stat.} {\bf 5}, 2493-2518 (2011).

\bibitem[Kvam et al.,(2012)]{kvam}
V.M. Kvam, P. Liu, Y. Si,
``A comparison of statistical methods for detecting differentially 
expressed genes from RNA-seq data",
{\sl Am. J. Botany} {\bf 99}, 248-256 (2012).

\bibitem[Li et al.,(2012)]{jli12}
J. Li, D.M. Witten, I.M. Johnsone, R. Tibshirani,
``Normalization, testing, and false discovery rate estimation for RNA-sequencing data",
{\sl Biostat.} {\bf 13}, 523-538 (2010).


\bibitem[Snedecor and Cochran,(1989)]{snedecor}
G.W. Snedecor and W.G. Cochran,
{\sl Statistical Methods}, eighth edition (Iowa State University Press: Ames, IW).
{\sl BMC Biotech.} {\bf 4}, 3 (1989).

\bibitem[Welsh,(1947)]{welsh}
B.L. Welsh,
``The generalization of 'Student's' problem when several different 
population variances are involved",
{\sl Biometrika} {\bf 34}, 28-35 (1947).

\bibitem[Cohen,(1988)]{cohenbook}
J. Cohen,
{\sl Statistical Power Analysis for the Behavioral Sciences},
2nd edition (Lawrence Erlbaum Associates, Inc. 1988).

\bibitem[Tu et al.,(2002)]{tu}
Y. Tu, G. Stolovitzky, U. Klein,
``Quantitative noise analysis for gene expression microarray experiments",
{\sl Proc. Nat. Acad. Sci. } {\bf 99}, 14031-14036 (2002).

\bibitem[Cui and Churchill,(2003)]{cui03}
X. Cui, G.A. Churchill,
``Statistical tests for differential expression in cDNA microarray 
experiments",
{\sl Genome Biol.} {\bf 4}, 210 (2003).

\bibitem[Jin et al.,(2001)]{gibson}
W. Jin,  R.M. Riley, R.D. Wolfinger, K.P. White, G. Passador-Gurgel, G. Gibson,
``The contributions of sex, genotype and age to transcriptional variance in Drosophila melanogaster",
{\sl Nat. Genet.} {\bf 29}, 389-395 (2001).

\bibitem[Alvord et al.,(2007)]{alvord}
W.G. Alvord, J.A. Roayaei, O.A. Qui\~{n}ones, K.T. Schneider,
``A microarray analysis for differential gene expression
in the soybean genome using Bioconductor and R",
{\sl Brief. Bioinf.} {\bf  8}, 1-13 (2007).

\bibitem[Pan,(2002)]{pan}
W. Pan, 
``A comparative review of statistical methods for discovering 
differentially expressed genes in replicated microarray experiments",
{\sl Bioinf.} {\bf 18}, 546-554 (2002).

\bibitem[Zhang and Cao,(2009)]{zhang}
S. Zhang, J. Cao,
``A close examination of double filtering with fold change and t test in 
microarray analysis",
{\sl BMC Bioinf.} {\bf 10}, 402 (2009).

\bibitem[Leek et al.,(2010)]{leek10}
J.T. Leek, R.B. Scharpf, H.C. Bravo, 
D. Simcha, B. Langmead, W.E. Johnson, D. Geman, K. Baggerly, R.A. Irizarry,
``Tackling the widespread and critical impact of batch effects in high-throughput data",
{\sl Nat. Rev. Genet.} {\bf 11}, 733-739 (2010).

\bibitem[Chen et al.,(2011)]{chen10}
Z. Chen, Q. Liu, M. McGee, M. Kong, X. Huang, Y. Deng,
``A gene selection method for GeneChip array data with small sample sizes",
{\sl BMC Genomics}, 12(suppl 5):S7 (2011).

\bibitem[Dozmorov and Lefkovits,(2009)]{igor}
I. Dozmorov, I. Lefkovits ,
``Internal standard-based analysis of microarray data. Part 1: analysis of
differential gene expressions",
{\sl Nucleic Acids Res.} {\bf 37}, 6323-6339 (2009).

\bibitem[Braga-Neto and Dougherty,(2004)]{braga}
U.M. Braga-Neto, E.R. Dougherty ,
``Is cross-validation valid for small-sample microarray classification?",
{\sl Bioinf.} {\bf 20}, 374-380 (2004).

\bibitem[Cui et al.,(2005)]{cui05}
X. Cui, J.T.G. Hwang, J. Qiu, N.J. Blades, G.A. Churchill,
``Improved statistical tests for differential gene expression by 
shrinking variance components estimates",
{\sl Biostat.} {\bf 6}, 59-75 (2005).

\bibitem[Wright and Simon,(2003)]{wright}
G.W. Wright, R.M. Simon,
``A random variance model for detection of differential gene expression in 
small microarray experiments",
{\sl Bioinf.} {\bf  19}, 2448-2455 (2003). 

\bibitem[Tusher et al.,(2001)]{sam}
V.G. Tusher, R. Tibshirani, G. Chu ,
``Significance analysis of microarrays applied to the ionizing
radiation response",
{\sl Proc. Nat. Acad. Sci. } {\bf 98}, 5116-5121 (2001).

\bibitem[Chu et al., (2007)]{sam-docu}
G. Chu, B. Narasimhan, R. Tibshirani, V. Tusher,
{\sl SAM: ``significance analysis of microarrays",
users guide and technical document}, v.3.0 (2007).
 
\bibitem[Schwender et al.,(2006)]{siggenes}
H. Schwender, A. Krause, K. Ickstadt,
``Identifying interesting genes with siggenes", 
{\sl RNews} {\bf 6}, 45-50 (2006).

\bibitem[Efron et al.,(2001)]{efron}
B. Efron, R. Tibshirani, J.D. Storey, V. Tusher,
``Empirical Bayes analysis of a microarray experiment",
{\sl J. Am. Stat. Assoc.} {\bf 96}, 1151-1160 (2001).

\bibitem[Wickham,(2009)]{ggplot2}
H. Wickham, {\sl ggplot2: Elegant Graphics for Data Analysis}
(Springer, 2009).

\bibitem[Gentleman et al.,(2004)]{biocond04}
R.C. Gentleman, V.J. Carey, D.M. Bates, et al.,
``Bioconductor: open software development for computational
biology and bioinformatics",
{\sl Genome Biol.} {\bf 5}, R80 (2004).

\bibitem[eds., Gentleman et al.,(2005)]{biocond}
eds.  R. Gentleman, W. Huber, V.J. Carey, R.A. Irizarry, S. Dudoit,
{\sl Bioinformatics and Computational Biology Solutions Using R and Bioconductor}
(Springer, 2005).

\bibitem[Yang et al.,(2005)]{deds}
Y.H. Yang, Y. Xiao, M.R. Segal,
``Identifying differentially expressed genes from microarray experiments via statistic synthesis",
{\sl Bioinf.} {\bf  21}, 1084-1093 (2005). 

\bibitem[Xiao et al.,(2012)]{pivalue}
Y. Xiao, T.H. Hsiao, U. Suresh, H.I.H. Chen, X. Wu, S.E. Wolf, Y. Chen,
``A novel significance score for gene selection and ranking",
{\sl Bioinf.}, in press (2012).

\bibitem[Ashburner et al.,(2000)]{go}
M. Ashburner, C.A. Ball, J.A. Blake, et al.,
``Gene Ontology: tool for the unification of biology"
{\sl Nat. Genet.} {\bf 25}, 25-29 (2000).

\bibitem[Li,(2006)]{li-bib}
W. Li,
``Three lectures on case-control genetic association analysis",
{\sl Brief. Bioinf.} {\bf 9}, 1-13 (2008).

\bibitem[Sirota et al.,(2009)]{sirota}
M. Sirota, M.A. Schaub, S. Batzoglou, W.H. Robinson, A.J. Butte,
``Autoimmune disease classification by inverse association with SNP alleles",
{\sl PLoS Genet.} {\bf 5}, e1000792 (2009).

\bibitem[Miclaus et al.,(2010)]{miclaus}
K Miclaus, M Chierici, C. Lambert, L. Zhang, S. Vega, H. Hong, 
S. Yin, C. Furlanello, R. Wolfinger, F. Goodsaid,
``Variability in GWAS analysis: the impact of genotype calling algorithm inconsistencies",
{\sl The Pharmacogenomics J.} {\bf 10}, 324-335 (2010).

\bibitem[Eisen et al.,(1998)]{eisen}
M.B. Eisen, P.T. Spellman, P.O. Brown, D. Botstein,
``Cluster analysis and display of genome-wide expression patterns",
{\sl Proc. Nat. Acad. Sci.} {\bf 95}, 14863-14868 (1998).

\bibitem[Yang et al.,(2002)]{ma}
Y.H. Yang, S. Dudoit, P. Luu, D.M. Lin, V. Peng, J. Ngal, T.P. Speed,
``Normalization for cDNA microarray data: a robust composite method 
addressing single and multiple slide systematic variation",
{\sl Nucleic Acids Res.} {\bf 30}, e15 (2002).

\bibitem[Benito et al.,(2004)]{pca}
M. Benito, J. Parker, Q. Du, J. Wu, D. Xiang, C.M. Perou, J.S. Marron,
``Adjustment of systematic microarray data biases",
{\sl Bioinf.} {\bf 20}, 105-114 (2004).

\bibitem[G\"{o}hlmann and Talloen,(2009)]{a4}
H. G\"{o}hlmann, W. Talloen,
{\sl Gene Expression Studies Using Affymetrix Microarrays}
(Chapman \& Hall/CRC, 2009).

\bibitem[Dozmorov and Centola,(2003)]{dga}
I. Dozmorov, M. Centola,
``An associative analysis of gene expression array data",
{\sl Bioinf.} {\bf 19}, 204-211 (2003).

\bibitem[Smyth,(2004)]{limma}
G.K. Smyth,
``Linear models and empirical Bayes methods for
assessing differential expression in microarray experiments",
{\sl Stat. Appl. Genet. Mol. Biol.} {\bf 3}, 3 (2004).

\bibitem[Wu et al.,(2003)]{maanova}
H. Wu, M. Kerr, X. Cui, G. Churchill,
``MAANOVA: a software package for the analysis of spotted
cDNA microarray experiments",
in {\sl The Analysis of Gene Expression Data Methods and Software}
eds. G Parmigiani, E Garrett, R Irizarry, S Zeger, pp.313-341 (Springer, 2003).

\bibitem[Dean and Raftery,(2005)]{nudge}
N. Dean, A.E. Raftery,
``Normal uniform mixture differential gene expression detection for cDNA microarrays",
{\sl BMC Bioinf.} {\bf 6}, 173 (2005).

\bibitem[Lin et al.,(2003)]{pickgene}
Y. Lin, S.T. Nadler, H. Lan, A.D. Attie, B.S. Yandell,
``Adaptive geen picking with microarray data: detecting important 
low abundance signals",
in {\sl The Analysis of Gene Expression Data Methods and Software}
eds. G Parmigiani, E Garrett, R Irizarry, S Zeger, pp.291-312 (Springer, 2003).

\bibitem[Pavelka et al.,(2004)]{plgem}
N. Pavelka, M. Pelizzola, C. Vizzardelli, M. Capozzoli, A. Splendiani, 
F. Granucci, P. Ricciardi-Castagnoli,
``A power law global error model for the identification
of differentially expressed genes in microarray data",
{\sl BMC Bioinf.} {\bf 5}, 203 (2004).

\bibitem[Cho et al., (2007)]{plpe}
H Cho, DM Smalley, MM Ross, D. Theodorescu, K. Ley, J.K. Lee,
``Statistical identification of differentially labelled peptides from
liquid chromatography tandem mass spectrometry",
{\sl Proteomics} {\bf 7}, 3681-3692 (2007).

\bibitem[\AA strand,(2008)]{plw}
M. \AA strand,
{\sl Normalization and Differential Gene Expression Analysis of
Microarray Data} (Ph.D Thesis, Department of Mathematics,
Chalmers University of Technology and G\"{o}teborg University, 2008).


\bibitem[Pearson et al.,(2009)]{puma}
R.D. Pearson, X. Liu, G. Sanguinetti, M. Milo, N.D. Lawrence, M. Rattray,
``puma: a Bioconductor package for propagating uncertainty in microarray analysis",
{\sl BMC Bioinf.} {\bf 10}, 211 (2009).

\bibitem[Breitling et al.,(2004)]{rankprod}
R. Breitling, P. Armengaul, A. Amtmann, P. Herzyk,
``Rank products: a simple, yet powerful new method to detect differentially 
regulated genes in replicated microarray experiments",
{\sl FEBS Lett.} {\bf 573}, 83-92 (2004).

\bibitem[Broberg,(2003)]{broberg}
P. Broberg,
``Statistical methods for ranking differentially expressed genes",
{\sl Genome Biol.} {\bf 4}, R41 (2004).

\bibitem[Scharpf et al.,(2009)]{xde}
R.B. Scharpf, H. Tjelmeland, G. Parmigiani, A.B. Nobel,
``A Bayesian model for cross-study differential gene expression",
{\sl J Am. Stat. Assoc. } {\bf 104}, 1295-1310 (2009).

\end{thebibliography}
\end{document}